\documentclass[12pt]{article}
\usepackage[top=25truemm,bottom=27truemm,left=22truemm,right=22truemm]{geometry}
\renewcommand{\baselinestretch}{1.15}
\usepackage{indentfirst}
\usepackage{graphicx}
\usepackage{float}
\usepackage{url}
\usepackage{array}
\usepackage{amsmath}
\usepackage{amssymb}
\usepackage{amsthm}
\usepackage{fancybox}
\usepackage{ascmac}
\usepackage{bm}
\usepackage{mathrsfs}
\usepackage{cancel}
\usepackage{physics}
\usepackage{color}
\usepackage{authblk}
\usepackage{appendix}
\usepackage{comment}
\usepackage{caption,latexsym,amsfonts,mathtools,here}
\usepackage{hyperref}
\usepackage{multirow}
\usepackage{tabularx}
\usepackage{booktabs}
\usepackage{tikz}
\usepackage{microtype}
\hypersetup{hidelinks,pdftitle={Entanglement, records, and quantum recovery in qubit chains},pdfauthor={Sugumi Kanno and Jiro Soda}}
\numberwithin{equation}{section}

\newcommand{\beqn}{\begin{eqnarray}}
\newcommand{\eeqn}{\end{eqnarray}}
\newcommand{\id}{\mathbf{1}}
\newcommand{\cA}{\mathcal{A}}

\newcommand{\cP}{\mathcal{P}}
\newcommand{\spanop}{\operatorname{span}}
\newtheorem{proposition}{Proposition}

\newcolumntype{Y}{>{\raggedright\arraybackslash}X}

\begin{document}
%==========================================================
\begin{titlepage}
\begin{flushright}
\hspace{7cm}{YITP-26-166, KOBE-COSMO-26-15}
% The unassigned YITP report-number placeholder in the supplied source was omitted.
\end{flushright}
\vspace{30pt}

\begin{center}
{\large{\textbf{Entanglement, records, and quantum recovery in qubit chains}}}

\vspace{25pt}

{Sugumi Kanno$^{1,2,3}$ and Jiro Soda$^4$}
\end{center}

\vspace{20pt}

\shortstack[l]
{\hspace{1.8cm}\it {\small $^1$Department of Physics, Kyushu University, Fukuoka, Japan} \\[5pt]
\hspace{1.8cm}\it {\small $^2$Quantum and Spacetime Research Institute, Kyushu University, Japan} \\[5pt]
\hspace{1.8cm}\it {\small $^3$Yukawa Institute for Theoretical Physics, Kyoto University, Japan} \\[5pt]
\hspace{1.8cm}\it {\small $^4$Department of Physics, Kobe University, Kobe 657-8501, Japan}}
\vspace{28pt}

%==========================================================
\begin{abstract}
 We investigate entanglement, readable records and quantum recovery using an exactly solvable chain of sequentially interacting qubits. For weak interactions, entanglement reaches a later environment site before that site carries an equally strong local record. The leading local record is encoded in an off-diagonal element of the reduced density matrix and can be detected by a Pauli measurement, while a population measurement gives a higher-order signal. We further characterize the accessibility of the relative phase through joint transverse correlations and derive the optimal fidelity for recovering an unknown quantum state from the system and an accessible environment fragment. We show that interactions acting only within the environment can redistribute readable records and change the region required for quantum recovery without affecting the reduced state of the system. 
\end{abstract}
\end{titlepage}
%==========================================================
%==========================================================

\begingroup
\renewcommand{\baselinestretch}{1.0}\small
\tableofcontents
\endgroup
\clearpage

%==========================================================
\section{Introduction}\label{sec:intro}

In a closed quantum system, entanglement with an environment does not by itself determine how quantum information is distributed among its subsystems. In particular, the entanglement of a subsystem with its complement, the presence of a locally readable record, and the possibility of recovering an unknown quantum state are distinct properties. Entanglement is characterized by the correlations across a bipartition, while a readable record depends on the distinguishability of conditional states within an accessible environment fragment. Quantum recovery requires, in addition, the preservation of the relative phase between the system branches. These distinctions become important when quantum correlations spread through an interacting environment.

The relation between entanglement and information transfer has been studied extensively in the context of decoherence and quantum Darwinism~\cite{Zurek2003,Ollivier2004,BlumeKohout2006,Zurek2009}. Decoherence is commonly characterized by the suppression of off-diagonal elements of a reduced density matrix, while quantum Darwinism concerns the formation of records of a preferred system observable in multiple environment fragments. Although entanglement plays an essential role in this process, the entanglement of an environment fragment with the rest of the system does not necessarily imply that the fragment contains a locally readable record ~\cite{Zwolak2010,Zwolak2014,Korbicz2014,Le2019}. Furthermore, the ability to distinguish the system
branch labels does not guaranty the recovery of
an arbitrary quantum superposition. These distinctions are related to accessible classical information and quantum correlations~\cite{Zwolak2013}, branching states and record networks~\cite{Touil2024,Ferte2024PRL,Ferte2024PRA}, and the no-broadcasting theorem~\cite{Barnum1996}. Quantum recovery provides a quantitative way to characterize the information accessible under specified controls ~\cite{Beny2007,Fawzi2015,Feller2023}.

In this paper, we investigate these issues using an exactly solvable chain of sequentially interacting qubits. A system qubit \(S\) first controls a rotation of an environment qubit \(E_1\), which subsequently controls a rotation of \(E_2\). The same interaction structure can be extended to an arbitrary number of environment qubits. Each environment qubit is initially prepared in \(|0\rangle\), so that the interaction with a later site becomes effective only after the preceding site has acquired a nonzero excitation amplitude. The dynamics is unitary and takes place in a finite-dimensional Hilbert space, allowing us to examine the redistribution of quantum information without assuming irreversible decoherence. The model is also related to collision models and structured spin environments ~\cite{Campbell2021,Ryan2021}.

We first compare the leading orders of entanglement generation and local record formation in the weak-interaction regime. For an environment qubit \(E_k\), let \(m_\star(E_k)\) denote the first nonvanishing order of the subleading Schmidt amplitude across the bipartition between \(E_k\) and the remaining qubits. We also introduce \(n_Z(E_k)\) as the leading order of the optimal distinguishability between the two system branches using measurements on \(E_k\) alone. For sequential interactions with pulse angles linear in a weak parameter, we obtain
\[
m_\star(E_k)=k,\qquad n_Z(E_k)=2k-1. \tag{1.1}
\]
The two orders coincide at the first environment site but differ at subsequent sites. This difference arises because the mixedness of an environment qubit can originate from entanglement within a fixed system branch, which does not necessarily provide an equally strong record of the system label. Thus, the onset of entanglement and the formation of a locally distinguishable record are characterized by different perturbative orders.

We further examine how the available measurements determine the accessibility of these records. The leading local record is encoded in an off-diagonal element of the reduced density matrix and can be detected by a single-site Pauli \(X\) measurement. In contrast, joint transverse correlations, involving operators such as \(X_S X_{E_1}\cdots X_{E_k}\) together with the corresponding \(Y_S\) quadrature, probe the relative phase between the system branches. These operators characterize different aspects of information accessibility and should be distinguished from the Pauli \(Y\) operators that generate the controlled rotations.

We then study the recovery of an unknown quantum state from the system and an accessible environment fragment. We derive the optimal recovery fidelity and show that interactions acting only within the environment can redistribute readable records and modify the degrees of freedom required for recovery, while leaving the reduced density matrix of the system unchanged. Consequently, the same local system state can correspond to different distributions of classical records and different requirements for recovering quantum coherence.

The organization of the paper is as follows. In Section 2, we introduce the quantities used to characterize entanglement, readable records, and quantum recovery. In Section 3, we define the sequential qubit model and the interaction protocols. In Section 4, we compare the leading orders of entanglement generation and local record formation. In Section 5, we derive the optimal recovery fidelity and examine its relation to records distributed outside the accessible region. Section 6 investigates the redistribution of phase information and its relation to multipartite entanglement. In Section 7, we generalize the results to an arbitrary chain of environment qubits. Section 8 summarizes our results. We set \(\hbar=1\) throughout.

\section{Entanglement, records, and quantum recovery}
\label{sec:entanglement-records-recovery}

We begin by introducing the quantities used to characterize entanglement,
locally readable records, and quantum recovery. These quantities depend on
different partitions of the total system and on the degrees of freedom
accessible to an observer. In particular, the entanglement of an environment
fragment with its complement does not necessarily imply that the fragment
contains a distinguishable record of the system state. Furthermore, the ability to distinguish the system
branch labels is different from the recovery of
an arbitrary quantum superposition. Table~\ref{tab:diagnostics} summarizes
the quantities used in the following analysis.

\subsection{Branches and fragments}
\label{subsec:branches-fragments}

Consider a system qubit $S$ with orthonormal basis states \begin{equation*}
|0\rangle =
\begin{pmatrix}
1\\
0
\end{pmatrix},
\qquad
|1\rangle =
\begin{pmatrix}
0\\
1
\end{pmatrix}.
\end{equation*} We use the Pauli operators
\begin{equation}
X=\begin{pmatrix}0&1\\1&0\end{pmatrix},\qquad
Y=\begin{pmatrix}0&-i\\i&0\end{pmatrix},\qquad
Z=\begin{pmatrix}1&0\\0&-1\end{pmatrix}.
\label{eq:paulidef}
\end{equation}
Subscripts specify the qubits on which the operators act, with identity
operators on the remaining qubits understood. The basis states are labeled by
$z=0,1$, with $Z|z\rangle=(-1)^z|z\rangle$. We refer to each component of the joint state
associated with a system basis label $z=0,1$
as a system branch. This terminology does not
imply a measurement or decoherence of the system.

\begin{table}[t]
\centering
\small
\setlength{\tabcolsep}{4pt}
\renewcommand{\arraystretch}{1.2}
\begin{tabular}{|p{0.17\linewidth}|p{0.23\linewidth}|p{0.22\linewidth}|p{0.28\linewidth}|}
\hline
Property & Quantity & Accessible subsystem or partition & Physical interpretation \\
\hline
Entanglement onset
& $P_A=1-\operatorname{Tr}\rho_A^2$, Schmidt amplitudes
& $A|\bar A$
& Whether $A$ and $\bar A$ are entangled \\
\hline
Readable branch record
& $D_Z(F)$
& Environment fragment $F$
& How well any measurement on $F$ can distinguish
the system branches $|0\rangle_S$ and $|1\rangle_S$ \\
\hline
Restricted record readout
& $D_Z^{\mathcal A_F}(F)$
& Observables in $\mathcal A_F$
& How well the allowed measurements on $F$ can
distinguish the system branches $|0\rangle_S$
and $|1\rangle_S$ \\
\hline
Quantum recovery
& $\eta_F=\|T_F\|_1$, $F_Q(F)$
& $S$ and an accessible fragment $F$
& How well the original logical qubit can
be recovered from $S$ and $F$ \\
\hline
Phase accessibility
& Joint $X/Y$ correlations
& Specified joint subsystem
& How well measurements can detect coherence
between the system branches \\
\hline
\end{tabular}
\caption{Quantities used to characterize entanglement, readable records,
and quantum recovery.}
\label{tab:diagnostics}
\end{table}

We consider a pure joint state of the form
\begin{equation}
|\Psi\rangle
=\sqrt{q}\,|0\rangle_S|r_0\rangle
+e^{i\phi}\sqrt{p}\,|1\rangle_S|r_1\rangle,
\qquad q=1-p,\quad 0<p<1.
\label{eq:branch-state}
\end{equation}
Here $p$ and $q$ are the branch probabilities, $\phi$ is their relative
phase, and $|r_z\rangle$ denotes the normalized environment state conditional
on the system label $z$. The conditional environment states need not be
orthogonal.

Let $E$ denote the full environment and $F\subseteq E$ an environment
fragment, with complement $\bar F=E\setminus F$. The corresponding conditional
reduced states are
\begin{equation}
\rho_F^{(z)}=\operatorname{Tr}_{\bar F}|r_z\rangle\langle r_z|.
\label{eq:conditional-fragment}
\end{equation}
The unconditional reduced state of the fragment is
$\rho_F=q\rho_F^{(0)}+p\rho_F^{(1)}$, while
$\rho_{SF}=\operatorname{Tr}_{\bar F}|\Psi\rangle\langle\Psi|$ retains the
system qubit. The complement $\bar F$ is taken within the environment and
therefore does not include $S$. For a general subsystem $A\subseteq SE$,
we instead use $\bar A$ for its complement in the full system and
$\rho_A=\operatorname{Tr}_{\bar A}|\Psi\rangle\langle\Psi|$.

\subsection{Entanglement across a cut}
\label{subsec:entanglement-cut}

We consider a bipartition $A|\bar A$ of the full
system, with the Hilbert space
$\mathcal H=\mathcal H_A\otimes\mathcal H_{\bar A}$.
For a globally pure state, entanglement across
this cut is characterized by the Schmidt
decomposition. When $A$ is a qubit, we write
\begin{equation}
|\Psi\rangle
=s_0|a_0\rangle|b_0\rangle+s_1|a_1\rangle|b_1\rangle,
\qquad s_0^2+s_1^2=1,\quad s_0\geq s_1\geq0,
\label{eq:schmidt-qubit}
\end{equation}
where the states $|a_i\rangle$ and $|b_i\rangle$ form orthonormal Schmidt
bases. The state is entangled across the cut if and only if $s_1\neq0$.

For an analytic family of states parameterized
by $\tau$ and initially product across this cut,
the contact order $m_\star(A)$ is defined as
the first nonvanishing order of a subleading
Schmidt amplitude~\cite{KannoSoda2026}. In particular, if
\begin{equation}
s_1(\tau)=a\tau^m+o(\tau^m),
\qquad a>0,
\label{contact-order}
\end{equation}
as $\tau\to0^+$, then $m_\star(A)=m$. For a subsystem with higher dimension,
the contact order is determined by the lowest nonvanishing order among its
subleading Schmidt amplitudes. If the state remains product in a neighborhood
of the origin, we define $m_\star(A)=\infty$.

We calculate the contact order using the purity deficit,
\begin{equation}
P_A=1-\operatorname{Tr}\rho_A^2.
\label{eq:purity-deficit}
\end{equation}
For a globally pure state, $P_A>0$ is equivalent
to entanglement across $A|\bar A$.
When $A$ is a qubit, $P_A=2\det\rho_A$,
and Eq.~\eqref{eq:schmidt-qubit} gives
\begin{equation}
P_A=2s_0^2s_1^2=2s_1^2(1-s_1^2)
=2a^2\tau^{2m}+o(\tau^{2m}).
\label{eq:purity-scaling}
\end{equation}
Thus, the leading order of the purity deficit is twice that of the subleading
Schmidt amplitude. This distinction is important when comparing entanglement
generation with the formation of locally readable records. The purity deficit
characterizes entanglement across a specified cut but does not, by itself,
imply irreversible decoherence or the presence of a distinguishable system
record.

\subsection{Locally readable branch records}
\label{subsec:branch-records}

To characterize the information accessible from an environment fragment $F$,
we consider the distinguishability of its two conditional states
$\rho_F^{(0)}$ and $\rho_F^{(1)}$. For equal prior probabilities, the optimal
distinguishability is given by their trace distance,
\begin{equation}
D_Z(F)=\frac12\left\|\rho_F^{(0)}-\rho_F^{(1)}\right\|_1,
\label{eq:record-distance}
\end{equation}
where $\|M\|_1=\operatorname{Tr}\sqrt{M^\dagger M}$ denotes the trace norm.
The subscript $Z$ specifies the system observable whose basis label is
recorded, rather than the measurement performed on the environment fragment.

The distinguishability satisfies $0\leq D_Z(F)\leq1$, with $D_Z(F)=1$
corresponding to perfectly distinguishable conditional states. For equal
priors, the optimal single-shot guessing probability is $[1+D_Z(F)]/2$.
We use equal priors to characterize the distinguishability independently
of the physical branch probabilities $p$ and $q$. For the physical priors,
the optimal guessing probability is instead
\begin{equation}
P_{\mathrm{guess}}^{(p,q)}(F)
=\frac12\left[1+\left\|q\rho_F^{(0)}-p\rho_F^{(1)}\right\|_1\right].
\label{eq:physical-priors}
\end{equation}
In this case, unequal prior probabilities can improve the guessing
probability even when the fragment contains no information about the
system label.

We define the record onset order $n_Z(F)$ as the first nonvanishing power in
\begin{equation}
D_Z(F)=d\tau^{n_Z}+o(\tau^{n_Z}),\qquad d>0.
\label{leading-weak-parameter}
\end{equation}
If the conditional states remain identical near $\tau=0$, we set
$n_Z(F)=\infty$. Like the contact order $m_\star$, the quantity $n_Z$
characterizes the leading weak-parameter dependence, rather than a finite
detection time. 
A higher onset order means that $D_Z(F)$
decreases faster as $\tau\to 0$.
The time required to distinguish the two system
branches experimentally also depends on the
measurement uncertainty and the chosen
detection threshold.

\subsection{Phase visibility and quantum recovery}
\label{subsec:visibility-recovery}

The distinguishability of the system branch labels
does not determine whether an unknown quantum
superposition can be recovered. In particular, a readable
record of the system label does not provide direct information about the
relative phase $\phi$. We therefore introduce a measure of coherence
accessible from the system $S$ and an environment fragment $F$.

Consider the dephasing map in the system $Z$ basis,
\begin{equation}
\Delta_Z(\rho)
=\sum_{z=0,1}
\bigl(|z\rangle\langle z|_S\otimes\mathbf{1}\bigr)
\rho
\bigl(|z\rangle\langle z|_S\otimes\mathbf{1}\bigr),
\label{eq:dephasing-map}
\end{equation}
which removes the off-diagonal blocks between the two system branches while
preserving their probabilities and conditional environment states. The
dephased state is used only as a reference for comparison and is not assumed
to arise from an additional physical process.

For joint access to $S$ and $F$, we define the phase visibility by
\begin{equation}
V(F)=\left\|\rho_{SF}-\Delta_Z(\rho_{SF})\right\|_1.
\label{eq:visibility}
\end{equation}
This quantity characterizes the distinguishability of the coherent state
from its dephased counterpart. For equal prior probabilities assigned to
these two hypotheses, the optimal guessing probability is $1/2+V(F)/4$.
For the state in Eq.~\eqref{eq:branch-state}, the visibility is bounded by
$2\sqrt{pq}$. The definition includes access to $S$, even when
$F=\varnothing$.

We next consider the recovery of an arbitrary input qubit,
including its coherence with a reference system.
Let $L$ denote the logical input qubit.
The conditional environment states define an encoding
isometry $W$ from $L$ into the joint system $SE$,
\begin{equation}
W|z\rangle_L=|z\rangle_S|r_z\rangle,
\label{eq:encoding}
\end{equation}
where $W^\dagger W=\mathbf{1}_L$ follows from the
orthogonality of the system basis states.
When only the system $S$ and an environment
fragment $F$ are accessible, the corresponding
quantum channel is
\begin{equation}
\mathcal N_F(\rho_L)
=\operatorname{Tr}_{\bar F}(W\rho_LW^\dagger).
\label{eq:fqdef}
\end{equation}

We define the optimal quantum recovery fidelity as
\begin{equation}
F_Q(F)
=\max_{\mathcal R_{SF\to L}}
F_e\left(\mathcal R_{SF\to L}\circ\mathcal N_F,
\frac{\mathbf{1}_L}{2}\right),
\label{eq:entfidelitydefinition}
\end{equation}
The maximization is taken over all completely positive trace-preserving recovery maps $\mathcal R_{SF\to L}$ from the accessible subsystem $SF$ to the logical qubit $L$.
The entanglement fidelity $F_e$ is defined by
\begin{equation}
\begin{aligned}
F_e\left(\Lambda,\frac{\mathbf{1}_L}{2}\right)
&=\langle\Phi|_{RL}
(\mathrm{id}_R\otimes\Lambda)
\bigl(|\Phi\rangle\langle\Phi|_{RL}\bigr)
|\Phi\rangle_{RL},\\
|\Phi\rangle_{RL}
&=\frac{|00\rangle_{RL}+|11\rangle_{RL}}{\sqrt{2}}.
\end{aligned}
\label{eq:entanglement-fidelity}
\end{equation}
Here $\mathrm{id}_R$ denotes the identity channel
on the reference qubit $R$, which remains
unaffected by the encoding and recovery operations. This definition tests
the recovery of an unknown quantum state, including its entanglement with
an external reference, rather than the recovery of the two basis states
alone. The maximally mixed input $\mathbf{1}_L/2$ specifies an unbiased
recovery test and is independent of the probabilities $p$ and $q$ in
Eq.~\eqref{eq:branch-state}. Perfect recovery corresponds to $F_Q(F)=1$.

The accessible region always includes the system $S$,
so $F=\varnothing$ corresponds to recovery from $S$
alone. An environment fragment can contain a readable
record of the system branch label without retaining
the relative phase of an arbitrary logical input.
Indeed, tracing out $S$ removes the off-diagonal
logical operators from the encoded state.
Quantum recovery therefore depends on the degrees
of freedom jointly accessible with $S$, rather than
on the record distinguishability of $F$ alone.

\subsection{Measurement restrictions and accessible coherence}
\label{subsec:restricted-observables}

The trace distance in Eq.~\eqref{eq:record-distance} assumes unrestricted
measurements on the environment fragment. In general, however, an observer
may have access only to a restricted set of observables. The distinguishability
of the conditional states then depends on the available measurements.

Let $\mathcal A_F\subseteq\mathcal B(\mathcal H_F)$
denote the algebra of accessible operators on the
environment fragment $F$, where
$\mathcal B(\mathcal H_F)$ is the space of linear
operators on $\mathcal H_F$.
We assume that $\mathcal A_F$ contains the identity
and is closed under Hermitian conjugation and
operator multiplication.
The corresponding restricted record
distinguishability is defined by
\begin{equation}
D_Z^{\mathcal A_F}(F)
=\frac12\sup_{\substack{O=O^\dagger\in\mathcal A_F\\\|O\|_\infty\leq1}}
\left|\operatorname{Tr}\left[
O\bigl(\rho_F^{(0)}-\rho_F^{(1)}\bigr)
\right]\right|,
\label{eq:restricted-record-distance}
\end{equation}
where $\|O\|_\infty$ is the operator norm. For unrestricted observables,
Eq.~\eqref{eq:restricted-record-distance} reduces to
Eq.~\eqref{eq:record-distance}. We denote the corresponding onset order by
$n_Z^{\mathcal A_F}(F)$, with the same infinite-order convention as for
unrestricted distinguishability.

This distinction is relevant to the two roles of transverse Pauli
measurements in our model. A local operator $X_{E_k}$ detects the
off-diagonal difference between the two conditional states of a single
environment qubit. By contrast, joint operators such as
$X_SX_{E_1}\cdots X_{E_k}$, together with the corresponding $Y_S$
quadrature, probe coherence between the global system branches. These are
readout observables; the controlled rotations in the model are generated
by Pauli $Y$ operators.

To characterize coherence under measurement restrictions,
let $\mathcal A_{\mathrm{obs}}$ denote the algebra
of observables accessible to a specified observer
on the joint system. We assume that
$\mathcal A_{\mathrm{obs}}$ contains the local
algebras $\mathcal A_F$ for the fragments under
consideration, extended by identity operators
on the remaining subsystems.
The corresponding phase visibility is defined by
\begin{equation}
V_{\mathcal A_{\mathrm{obs}}}(\rho)
=\sup_{\substack{
O=O^\dagger\in\mathcal A_{\mathrm{obs}}\\
\|O\|_\infty\leq1}}
\left|\operatorname{Tr}\left[
O\bigl(\rho-\Delta_Z(\rho)\bigr)
\right]\right|.
\label{eq:restricted-visibility}
\end{equation}

Let $\mathcal F_{\mathrm{obs}}$ be a nonempty collection of nonempty
environment fragments. For $0\leq\delta<1$ and $\epsilon\geq0$, an
operational condition for locally readable records with unobservable
coherence is
\begin{equation}
\min_{F\in\mathcal F_{\mathrm{obs}}}
D_Z^{\mathcal A_F}(F)\geq1-\delta,
\qquad V_{\mathcal A_{\mathrm{obs}}}(\rho)\leq\epsilon.
\label{eq:accessible-record-condition}
\end{equation}
The first inequality requires the equal-prior branch-guessing error
to be at most $\delta/2$ for every selected fragment, while the second
bounds the observable distinction between the coherent and dephased
states.

Equation~\eqref{eq:accessible-record-condition} characterizes the
accessibility of branch records under specified
measurement restrictions, but does not by itself
establish their objectivity. In particular, spectrum
broadcast structure requires a branch-diagonal state with conditionally
independent records in distinct fragments and mutually orthogonal record
states for different labels [7, 8]. These additional conditions are not
imposed by Eq.~\eqref{eq:accessible-record-condition}. Moreover, when
observers cannot combine their measurement outcomes, joint products of
their observables must not be included among the accessible measurements.

The distinction between unrestricted and restricted measurements can be
seen directly for the states $|0\rangle$ and $|1\rangle$. Although their
unrestricted distinguishability is $D_Z=1$, an observer restricted to
the algebra $\operatorname{span}\{\mathbf{1},X\}$ obtains identical
measurement statistics and hence $D_Z^{\mathcal A}=0$. Thus, record
distinguishability and phase visibility must be compared using measurement
resources appropriate to the same observer.

Measurement restrictions do not necessarily determine the quantum operations
available for recovery. If $\mathcal C_F$ denotes the allowed recovery maps
acting on $SF$, we define
\begin{equation}
F_Q^{\mathcal C}(F)
=\sup_{\mathcal R\in\mathcal C_F}
F_e\left(\mathcal R\circ\mathcal N_F,
\frac{\mathbf{1}_L}{2}\right)
\leq F_Q(F).
\label{eq:restricted-recovery}
\end{equation}
The unrestricted recovery fidelity therefore provides an upper bound
when the available controls are limited. Relations between unrestricted
phase visibility and recovery fidelity need not remain valid when
different restrictions are imposed on measurements and recovery
operations.

Finally, for a unitary transformation $U_E$ acting only on the environment,
\begin{equation}
\operatorname{Tr}_E\left[
(\mathbf{1}_S\otimes U_E)\rho
(\mathbf{1}_S\otimes U_E^\dagger)
\right]=\rho_S.
\label{eq:environment-only-unitary}
\end{equation}
Thus, environment-only interactions leave the reduced state of the system
unchanged, although the distribution of locally readable records and the
degrees of freedom required for quantum recovery can change. This property
will play a central role in the following analysis.

\section{Sequential qubit model}
\label{sec:sequential-qubit-model}

\subsection{Sequential controlled rotations}
\label{subsec:sequential-rotations}

We consider three qubits consisting of a system $S$
and two environment qubits $E_1$ and $E_2$.
The system interacts directly with $E_1$, while
$E_2$ interacts only with $E_1$.
Initially, the system qubit is prepared in a coherent
superposition of the basis states $|0\rangle_S$
and $|1\rangle_S$, while both environment qubits
are in the state $|0\rangle$.
The initial state is therefore
\begin{equation}
|\Psi_0\rangle=
\left(\sqrt{q}\,|0\rangle_S+e^{i\phi}\sqrt{p}\,|1\rangle_S\right)
|0\rangle_{E_1}|0\rangle_{E_2}, \qquad q=1-p,\quad 0<p<1.
\label{eq:initial-three-qubit-state}
\end{equation}
We apply two controlled rotations in sequence.
The initial state $|0\rangle$ of each environment
qubit is referred to as the ready state.
This is a preparation condition and does not
require $|0\rangle$ to be the ground state of
a specified free Hamiltonian.
With the Pauli convention of Eq.~\eqref{eq:paulidef},
the rotation about the $y$ axis is
\begin{equation}
R_y(\alpha)=e^{-i\alpha Y/2},\qquad
R_y(\alpha)|0\rangle
=\cos\frac{\alpha}{2}\,|0\rangle
+\sin\frac{\alpha}{2}\,|1\rangle.
\label{eq:y-rotation}
\end{equation}

To implement controlled rotations, we couple
the Pauli $Y$ operator of the target qubit
to the projector $|1\rangle\langle1|$
of the control qubit. The corresponding
Hamiltonians are
\begin{equation}
H_1=g|1\rangle\langle1|_S\otimes Y_{E_1},
\qquad
H_2=\lambda|1\rangle\langle1|_{E_1}\otimes Y_{E_2},
\label{eq:controlled-hamiltonians}
\end{equation}
where $g$ and $\lambda$ are real coupling strengths
and identity operators on spectator qubits are omitted.
For pulse durations $t_1$ and $t_2$, the corresponding
time-evolution operators are
\begin{equation*}
U_1(\theta)=e^{-iH_1t_1},
\qquad
U_2(\chi)=e^{-iH_2t_2},
\qquad
\theta=2gt_1,\quad \chi=2\lambda t_2.
\end{equation*}
Using Eq.~\eqref{eq:y-rotation}, we obtain
\begin{equation}
U_1(\theta)
=|0\rangle\langle0|_S\otimes\mathbf{1}_{E_1}
+|1\rangle\langle1|_S\otimes R_y^{E_1}(\theta),
\label{eq:first-controlled-rotation}
\end{equation}
and
\begin{equation}
U_2(\chi)
=|0\rangle\langle0|_{E_1}\otimes\mathbf{1}_{E_2}
+|1\rangle\langle1|_{E_1}\otimes R_y^{E_2}(\chi).
\label{eq:second-controlled-rotation}
\end{equation}
We apply the first pulse before the second,
so that the total evolution operator is
$U(\theta,\chi)=U_2(\chi)U_1(\theta)$.
The ordering is important because the second
interaction acts nontrivially only after the first
has generated an $E_1=1$ component.

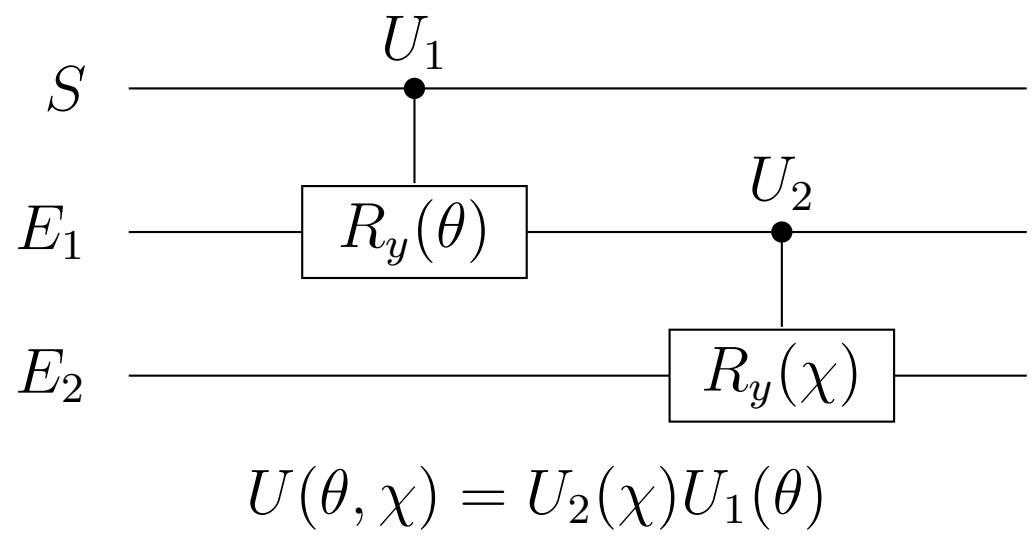
\begin{figure}[htbp]
\centering
\begin{tikzpicture}[x=1.35cm,y=0.95cm]
\foreach \y/\name in {1/S,0/{E_1},-1/{E_2}} {
  \draw (0.5,\y)--(4.9,\y);
  \node[left] at (0.4,\y) {$\name$};
}
\draw (1.9,1)--(1.9,0.34);
\fill (1.9,1) circle (2pt);
\draw[fill=white] (1.35,-0.32) rectangle (2.45,0.32);
\node at (1.9,0) {$R_y(\theta)$};
\draw (3.7,0)--(3.7,-0.66);
\fill (3.7,0) circle (2pt);
\draw[fill=white] (3.15,-1.32) rectangle (4.25,-0.68);
\node at (3.7,-1) {$R_y(\chi)$};
\node at (1.9,1.32) {$U_1$};
\node at (3.7,0.34) {$U_2$};
\node at (2.5,-1.85) {$U(\theta,\chi)=U_2(\chi)U_1(\theta)$};
\end{tikzpicture}
\caption{Sequential controlled rotations. The first gate couples $S$ to $E_1$, and the second couples $E_1$ to $E_2$. Filled dots denote controls, and boxes denote conditional rotations. The horizontal direction indicates gate order, rather than elapsed time. Both environment qubits start in $\ket0$.}
\label{fig:circuit}
\end{figure}

For convenience, we define
$c_\theta=\cos(\theta/2)$, $s_\theta=\sin(\theta/2)$,
$c_\chi=\cos(\chi/2)$, and $s_\chi=\sin(\chi/2)$.
We order the basis states as $S,E_1,E_2$, so that
$|110\rangle=|1\rangle_S|1\rangle_{E_1}|0\rangle_{E_2}$.
The final state is
\begin{equation}
\begin{aligned}
|\Psi(\theta,\chi)\rangle
={}&\sqrt{q}\,|000\rangle
+e^{i\phi}\sqrt{p}\,c_\theta|100\rangle
+e^{i\phi}\sqrt{p}\,s_\theta
\bigl(c_\chi|110\rangle+s_\chi|111\rangle\bigr).
\end{aligned}
\label{eq:sequential-final-state}
\end{equation}
Using the notation of Eq.~(\ref{eq:branch-state}), the conditional
environment states are
\begin{equation}
|r_0\rangle=|00\rangle,\qquad
|r_1\rangle=c_\theta|00\rangle
+s_\theta c_\chi|10\rangle
+s_\theta s_\chi|11\rangle.
\label{eq:sequential-conditional-states}
\end{equation}
The absence of a $|01\rangle$ component reflects the interaction order:
$E_2$ can be rotated only when $E_1$ is in $|1\rangle$.

\subsection{Interaction protocols}
\label{subsec:interaction-protocols}

To examine the weak-interaction behavior of the sequential circuit,
we take equal pulse durations $t_1=t_2=\tau$. The resulting
family of pulse angles is
\begin{equation}
\mathcal P_{\mathrm{on}}:\qquad
\theta=2g\tau,\qquad \chi=2\lambda\tau,
\qquad \tau\to0^+.
\label{eq:weak-pulse-protocol}
\end{equation}
Here $\tau$ is the duration of each pulse, so the total duration of the
two-pulse sequence is $2\tau$. Varying $\tau$ compares the final states
of different sequential-pulse experiments, rather than the evolution
under a single fixed Hamiltonian. Expanding Eq.~\eqref{eq:sequential-final-state} we obtain
\begin{equation}
\begin{aligned}
|\Psi(\tau)\rangle
={}&\sqrt{q}\,|000\rangle
+e^{i\phi}\sqrt{p}\,\Biggl[
\left(1-\frac{g^2\tau^2}{2}\right)|100\rangle
+g\tau|110\rangle
+g\lambda\tau^2|111\rangle\Biggr]+O(\tau^3).
\end{aligned}
\label{eq:weak-pulse-state}
\end{equation}
The $|110\rangle$ and $|111\rangle$ amplitudes begin at orders $\tau$
and $\tau^2$, respectively, reflecting the one- and two-step
interaction processes. The corresponding entanglement and record orders
are evaluated in Section~4.

For comparison, we consider continuous evolution under a fixed
Hamiltonian with both interactions in Eq.~(\ref{eq:controlled-hamiltonians}) present simultaneously,
\begin{equation}
H_{\mathrm{on}}=H_1+H_2,\qquad
|\Psi_{\mathrm{on}}(t)\rangle
=e^{-iH_{\mathrm{on}}t}|\Psi_0\rangle.
\label{eq:simultaneous-hamiltonian}
\end{equation}
Here $t$ denotes elapsed physical time. Since $H_1$ and $H_2$ do not
commute, this evolution is different from the sequential-pulse
protocol. The $S=0$ branch remains unchanged, while the
$S=1$ branch evolves under the conditional
environment Hamiltonian
\begin{equation}
H^{(1)}=gY_{E_1}\otimes\mathbf{1}_{E_2}
+\lambda|1\rangle\langle1|_{E_1}\otimes Y_{E_2}.
\label{eq:conditionalh}
\end{equation}
The superscript $(1)$ denotes the system branch
label $z=1$. We denote the corresponding
conditional environment state under the
simultaneous-interaction evolution by
$|r_1^{\mathrm{on}}(t)\rangle
=e^{-iH^{(1)}t}|00\rangle$,
where ``$\mathrm{on}$" distinguishes this evolution
from the sequential-pulse protocol.
Expanding through third order in $t$ using
Eq.~(\ref{eq:hpowers}), we obtain
\begin{equation}
\begin{aligned}
|r_1^{\mathrm{on}}(t)\rangle
={}&\left(1-\frac{g^2t^2}{2}\right)|00\rangle
-\frac{g^2\lambda t^3}{6}|01\rangle
+\left[gt-\frac{g(g^2+\lambda^2)t^3}{6}\right]|10\rangle
+\frac{g\lambda t^2}{2}|11\rangle+O(t^4).
\end{aligned}
\label{eq:fixed-hamiltonian-state}
\end{equation}
Unlike the sequential circuit, the simultaneous evolution generates
a $|01\rangle$ component at third order, because the first interaction
can return $E_1$ to $|0\rangle$ after $E_2$ has been excited. 
This term contributes to the off-diagonal element
of the conditional reduced state $\rho_{E_2}^{(1)}$
and must be retained when calculating
$D_Z^{\rm on}(E_2;t)$.

We finally consider the redistribution of an established record within
the environment. Setting $\theta=\pi$ makes the system branch labels
perfectly distinguishable by measuring $E_1$. We then vary only the
second pulse along
\begin{equation}
\mathcal P_{\mathrm{fan}}:\qquad
\theta=\pi,\qquad 0\leq\chi\leq\pi.
\label{eq:fanout-protocol}
\end{equation}
As $\chi$ increases, the record becomes accessible from $E_2$ as well.
At $\chi=\pi$, either environment qubit provides a perfect record of
the branch label. This copies the chosen basis label, not an arbitrary
unknown quantum state. Because the second pulse acts only within the
environment, the reduced state $\rho_S$ remains unchanged, as follows
from Eq.~\eqref{eq:environment-only-unitary}. Thus the distribution of
readable records can change without any change in the reduced state of
$S$.

\section{Onset of entanglement and readable records}\label{sec:onset}
\subsection{Purity deficits and record distinguishability}

For the pure state in Eq.~\eqref{eq:sequential-final-state},
the purity deficits of $S$, $E_1$, and $E_2$
characterize entanglement across $S|E_1E_2$,
$E_1|SE_2$, and $E_2|SE_1$, respectively.
Using the partial-trace relation in Eq.~\eqref{eq:partialtracecomponents},
the reduced state of $S$ is
\begin{equation}
\rho_S=
\begin{pmatrix}
q & \sqrt{pq}\,e^{-i\phi}c_\theta\\
\sqrt{pq}\,e^{i\phi}c_\theta & p
\end{pmatrix}.
\label{eq:rhos}
\end{equation}
Using the purity deficit defined in
Eq.~\eqref{eq:purity-deficit} and the determinants in
Eq.~\eqref{eq:determinants}, we obtain
\begin{equation}
P_S=2pq\,s_\theta^2.
\label{eq:ps}
\end{equation}
Similarly, the purity deficits of the two
environment qubits are
\begin{align}
P_{E_1}
&=2p\,s_\theta^2
\bigl(q+p c_\theta^2s_\chi^2\bigr),\\
P_{E_2}
&=2p\,s_\theta^2
\bigl(1-p s_\theta^2\bigr)s_\chi^2.
\label{eq:pe2}
\end{align}

The off-diagonal element of $\rho_S$ is
proportional to the overlap
$\langle r_0|r_1\rangle=c_\theta$
of Eq.~\eqref{eq:sequential-conditional-states}.
Since the second pulse acts only on the
environment, this overlap is independent
of $\chi$. In particular, at $\theta=\pi$,
\begin{equation}
\rho_S=q|0\rangle\langle0|
+p|1\rangle\langle1|,
\qquad
P_S=2pq\quad(\theta=\pi).
\end{equation}
Thus, the reduced state of $S$ remains
unchanged as $\chi$ varies, even though
the distribution of records within the
environment can change.

To determine which fragments carry a readable
system-branch record, we evaluate the
distinguishability $D_Z(F)$ defined in Eq.~\eqref{eq:record-distance}.
Using the conditional environment states in
Eq.~\eqref{eq:sequential-conditional-states} and the trace-distance
calculation in Eqs.~\eqref{eq:conditionalmatrices} and
\eqref{eq:tracelessmatrix}, we obtain
\begin{align}
D_Z(E_1E_2)
&=|s_\theta|,\\
D_Z(E_1)
&=|s_\theta|
\sqrt{1-c_\theta^2s_\chi^2},\label{eq:dz1}\\
D_Z(E_2)
&=s_\theta^2|s_\chi|\label{eq:dz2}.
\end{align}
Here $E_1E_2$ denotes joint access to both
environment qubits. The joint distinguishability
is independent of $\chi$, whereas the individual
distinguishabilities can vary with the second
rotation. These expressions assume arbitrary
measurements on the indicated fragments.
In particular, the off-diagonal elements
of $\rho_{E_2}^{(1)}$ in
Eq.~\eqref{eq:conditionalmatrices}
contribute to $D_Z(E_2)$, so the
distinguishability cannot be determined
from the $|1\rangle$ populations alone.

\subsection{Entanglement and record onset orders at $E_2$}

We first compare when $E_2$ becomes entangled with the rest of the system and when measurements on $E_2$ begin to distinguish the two system branches. Substituting the weak-pulse protocol in Eq.~\eqref{eq:weak-pulse-protocol} into Eqs.~\eqref{eq:ps}--\eqref{eq:pe2} and expanding in powers of $\tau$, we obtain
\begin{align}
 P_S&=2pqg^2\tau^2+O(\tau^4),\label{eq:psweak}\\
 P_{E_1}&=2pqg^2\tau^2+O(\tau^4),\label{eq:pe1weak}\\
 P_{E_2}&=2pg^2\lambda^2\tau^4+O(\tau^6).
 \label{eq:pe2weak}
\end{align}
We assume $0<p<1$ and $g\lambda\ne 0$. The factor $q=1-p$ in $P_{E_1}$ comes from the superposition of the two system states. If $p=1$, only $\ket1_S$ is present, and the first pulse changes $E_1$ without entangling it with $S$. Using the definition of the contact order in Eq.~\eqref{contact-order} and its relation to the purity deficit in Eq.~\eqref{eq:purity-scaling}, we obtain
\begin{equation}
 (m_\star(S),m_\star(E_1),m_\star(E_2))=(1,1,2).
 \label{eq:contactprofile}
\end{equation}
For $E_2$, the purity deficit starts at order $\tau^4$, so the subleading Schmidt amplitude across $E_2|SE_1$ starts at order $\tau^2$. This is consistent with the two-step interaction: $E_2$ is affected only after the first pulse has generated an $E_1=1$ component.

On the other hand, expanding $D_Z(E_2)$ in Eq.~\eqref{eq:dz2} and using the definition of the record onset order in Eq.~\eqref{leading-weak-parameter}, we find
\begin{equation}
 D_Z(E_2)=g^2\abs{\lambda}\,\tau^3+O(\tau^5),
 \qquad n_Z(E_2)=3.
 \label{eq:recordweak}
\end{equation}
Thus, entanglement across $E_2|SE_1$ begins at second order in the Schmidt amplitude, whereas the ability to distinguish the system branches by measuring $E_2$ alone begins at third order. These orders describe the behavior as $\tau\to0$, not a finite waiting time: both quantities are nonzero for arbitrarily small positive $\tau$. Entanglement across $E_2|SE_1$ should also not be confused with pairwise entanglement between $S$ and $E_2$. The reduced state of $SE_2$ is separable in this model, as shown in Section~\ref{sec:spectrum}.

To see why the two onset orders differ, consider the environment state conditioned on $S=1$. In this branch, $E_1$ and $E_2$ can be entangled even though the value of $S$ is known. The reduced state $\rho_{E_2}^{(1)}$ can therefore be mixed without providing a distinction between the two system branches. To separate these contributions, let $\mathcal P(\sigma)=1-\Tr\sigma^2$. Since $\rho_{E_2}=q\ket0\bra0+p\rho_{E_2}^{(1)}$, the identity in Eq.~\eqref{eq:variancegeneral} gives
\begin{equation}
 P_{E_2}=p\,\mathcal P(\rho_{E_2}^{(1)})
         +2pq\,D_Z(E_2)^2.
 \label{eq:conditionalvariance}
\end{equation}
The first term measures the mixedness of $E_2$ within the $S=1$ branch, which comes from entanglement between $E_1$ and $E_2$. The second term depends on how well measurements on $E_2$ distinguish the two system branches. For the sequential circuit, Eq.~\eqref{eq:conditionalmatrices} gives
\begin{equation}
 \mathcal P(\rho_{E_2}^{(1)})
 =2s_\theta^2c_\theta^2s_\chi^2.
 \label{eq:conditionalpurity}
\end{equation}
In the weak-pulse limit, the first term of Eq.~\eqref{eq:conditionalvariance} starts at order $\tau^4$, while the second starts at order $\tau^6$. The leading purity deficit of $E_2$ therefore comes from entanglement within the $S=1$ branch, rather than from the distinguishability of the system branches.

The different orders can also be seen directly from the amplitudes in Eq.~\eqref{eq:sequential-conditional-states}. In the $S=1$ branch, the amplitudes of $\ket{00}$ and $\ket{11}$ begin at orders $\tau^0$ and $\tau^2$, respectively, and give entanglement at second order. By contrast, the off-diagonal element of $\rho_{E_2}^{(1)}$ comes from the amplitudes of $\ket{10}$ and $\ket{11}$, which begin at orders $\tau$ and $\tau^2$. Their product is of order $\tau^3$. The partial trace over $E_1$ removes the interference between $\ket{00}$ and $\ket{11}$ because they have different $E_1$ labels. This explains why $D_Z(E_2)$ starts one order later than the subleading Schmidt amplitude.

\subsection{Onset orders under a fixed Hamiltonian}

In Section~4.2, we obtained the onset orders from the weak-pulse protocol of Eq.~\eqref{eq:weak-pulse-protocol}. We now examine whether the same orders arise during continuous evolution under the fixed Hamiltonian $H_{\mathrm{on}}=H_1+H_2$ in Eq.~\eqref{eq:simultaneous-hamiltonian}. Unlike the sequential-pulse protocol, both interactions act simultaneously, and $t$ is the elapsed physical time.

Using the conditional environment state in Eq.~\eqref{eq:fixed-hamiltonian-state} and the calculations in Eqs.~\eqref{eq:fixedB}--\eqref{eq:fixedDproof}, we obtain
\begin{align}
 P_S(t)&=2pqg^2t^2+O(t^4),
 \label{eq:fixedps}\\
 P_{E_1}(t)&=2pqg^2t^2+O(t^4),
 \label{eq:fixedpe1}\\
 P_{E_2}(t)&=\frac12pg^2\lambda^2t^4+O(t^6),
 \label{eq:fixedpe2}\\
 D_Z^{\rm on}(E_2;t)
 &=\frac{g^2\abs{\lambda}}{3}t^3+O(t^5).
 \label{eq:fixedrecord}
\end{align}
Equation~\eqref{eq:fixedpe2}, together with the relation between the purity deficit and the Schmidt amplitude in Eq.~\eqref{eq:purity-scaling}, gives $m_\star(E_2)=2$. Similarly, Eq.~\eqref{eq:fixedrecord} and the definition in Eq.~\eqref{leading-weak-parameter} give $n_Z(E_2)=3$. The two onset orders are therefore the same as those found for the sequential-pulse protocol, although their coefficients differ.

The difference in the coefficients can be understood from the conditional reduced state of $E_2$. Under the fixed Hamiltonian, the off-diagonal element of $\rho_{E_2}^{(1)}$ receives contributions from both the $\ket{00}$--$\ket{01}$ and the $\ket{10}$--$\ket{11}$ interference terms. Using Eq.~\eqref{eq:fixedu}, we find
\begin{equation}
 (\rho_{E_2}^{(1)})_{01}
 =-\frac{g^2\lambda t^3}{6}
   +\frac{g^2\lambda t^3}{2}+O(t^5)
 =\frac{g^2\lambda t^3}{3}+O(t^5).
 \label{eq:fixedoffdiag}
\end{equation}
The $\ket{01}$ amplitude is absent in the sequential-pulse protocol, but appears at third order under the fixed Hamiltonian because $E_1$ can return to $\ket0$ after exciting $E_2$. Consequently, the coefficient of $D_Z^{\rm on}(E_2;t)$ cannot be obtained from the $\ket{11}$ amplitude alone. The two protocols have the same onset orders, but different interference contributions determine their coefficients.

\subsection{Effects of initial preparation and measurement choice}
\label{sec:assumptions}

We next examine which assumptions determine the onset orders at $E_2$. For the original preparation, with both environment qubits initially in $\ket0$, changing nonzero $g$ and $\lambda$ or a fixed $0<p<1$ changes the coefficients but not the orders. We first change the initial state of $E_1$ and then consider different measurements on $E_2$.

Let $E_1$ initially be in $\ket+=(\ket0+\ket1)/\sqrt2$, while $E_2$ remains in $\ket0$. We apply the controlled rotations in Eqs.~\eqref{eq:first-controlled-rotation} and \eqref{eq:second-controlled-rotation}. Since the initial state of $E_1$ already contains a $\ket1$ component, the second controlled rotation can change $E_2$ even when $S=0$. From Eqs.~\eqref{eq:misalignedprobabilities} and \eqref{eq:misalignedconditional}, we obtain
\begin{equation}
 D_Z^{(+)}(E_2)=\frac12\abs{\sin\theta\,s_\chi}
             =\abs{g\lambda}\tau^2+O(\tau^4).
 \label{eq:misalignedrecord}
\end{equation}
Let $a$ be the probability that $E_1$ is in $\ket1$ immediately before the second controlled rotation. Equation~\eqref{eq:misalignedprobabilities} gives $a=qa_0+pa_1=(1+p\sin\theta)/2$. Using the conditional states in Eq.~\eqref{eq:misalignedconditional}, we obtain the purity deficit of $E_2$ as
\begin{equation}
 P_{E_2}^{(+)}=2a(1-a)s_\chi^2
              =\frac{\lambda^2}{2}\tau^2+O(\tau^4).
 \label{eq:misalignedpurity}
\end{equation}
Equations~\eqref{contact-order}, \eqref{eq:purity-scaling}, and \eqref{leading-weak-parameter} then give $(m_\star(E_2),n_Z(E_2))=(1,2)$, rather than $(2,3)$ as in Section~4.2. Thus, the onset orders depend on the initial preparation even when the two controlled rotations are unchanged.

We now return to the original preparation and ask how the measured observable affects the distinction between the system branches. From the conditional reduced states in Eq.~\eqref{eq:conditionalmatrices}, their difference is
\begin{equation}
 \Delta_2\equiv\rho_{E_2}^{(0)}-\rho_{E_2}^{(1)}
 =s_\theta^2s_\chi^2 Z_{E_2}
  -s_\theta^2c_\chi s_\chi X_{E_2}.
 \label{eq:delta2XZ}
\end{equation}
The $Z_{E_2}$ term describes a population difference, while the $X_{E_2}$ term comes from the off-diagonal elements of the conditional density matrices. Which term is accessible depends on the measurements performed on $E_2$.

If only $Z_{E_2}$ is measured, the observable algebra is $\cA_z=\spanop\{\id,Z_{E_2}\}$. Using the restricted distinguishability in Eq.~\eqref{eq:restricted-record-distance} and the difference in Eq.~\eqref{eq:delta2XZ}, we obtain
\begin{equation}
 D_Z^{\cA_z}(E_2)=s_\theta^2s_\chi^2
                  =g^2\lambda^2\tau^4+O(\tau^6).
 \label{eq:zrestrictedsecond}
\end{equation}
Thus, a population measurement distinguishes the two system branches only at fourth order in $\tau$. By contrast, measuring $X_{E_2}$ selects the off-diagonal term of Eq.~\eqref{eq:delta2XZ}. For $\cA_x=\spanop\{\id,X_{E_2}\}$, the same calculation gives
\begin{equation}
 D_Z^{\cA_x}(E_2)=s_\theta^2\abs{c_\chi s_\chi}
                  =g^2\abs{\lambda}\,\tau^3+O(\tau^5).
 \label{eq:xrestrictedsecond}
\end{equation}
This has the same leading power and coefficient as the unrestricted distinguishability $D_Z(E_2)$ in Eq.~\eqref{eq:recordweak}. A fixed $X_{E_2}$ measurement is therefore sufficient at leading order in the weak-pulse limit; the measurement direction need not depend on the pulse angles. It distinguishes the system branches, but does not determine their relative phase $\phi$. Joint $X/Y$ correlations involving $S$ are considered in Section~\ref{sec:spectrum}, and the optimal local measurement direction at finite pulse angles is given in Section~\ref{sec:chainonset}.

For the original preparation, the three onset orders are
\begin{equation}
 m_\star(E_2)=2,\qquad n_Z^{\cA_x}(E_2)=3,
 \qquad n_Z^{\cA_z}(E_2)=4.
 \label{eq:threeordersE2}
\end{equation}
The first describes entanglement across $E_2|SE_1$. The other two describe the distinguishability of the system branches under different measurement restrictions. Thus, even for the same quantum state, the leading observable distinction can depend on the available measurement. Section~\ref{sec:chainonset} extends this result to a longer chain and to pulse angles with different leading powers.
\label{par:chainpreview}

\section{Quantum recovery and readable records}
\subsection{Quantum recovery from $S$ and an environment fragment}

In Section~4, we asked how well measurements on an environment fragment distinguish the system branches. We now consider a different question: can an unknown quantum state be recovered when only $S$ and a fragment $F$ of the environment are accessible? Distinguishing the branch labels does not by itself guarantee the recovery of their relative coherence.

The encoding in Eq.~\eqref{eq:encoding} associates each system basis state with a conditional environment state $\ket{r_z}$. After the inaccessible environment $\bar F$ is traced out, the operator connecting the two branches is
\begin{equation}
 T_F=\Tr_{\bar F}\ket{r_0}\bra{r_1},\qquad
 \eta_F=\norm{T_F}_1.
 \label{eq:tf}
\end{equation}
The operator $T_F$ determines the off-diagonal system block of $\rho_{SF}$; it is not generally a density matrix. Its trace norm $\eta_F$, which satisfies $0\leq\eta_F\leq1$, characterizes the coherence available for recovery from $SF$. We call $\eta_F$ the coherent recovery amplitude. This differs from $D_Z(F)$, which describes the distinguishability of the branch labels from measurements on $F$ alone.

For the encoding in Eq.~\eqref{eq:encoding} and the optimal recovery fidelity defined in Eq.~\eqref{eq:entfidelitydefinition}, we obtain
\begin{equation}
 F_Q(F)=\frac{1+\eta_F}{2}.
 \label{eq:optimalrecovery}
\end{equation}
Appendix~\ref{app:recovery} constructs a deterministic recovery operation that achieves this result and proves that no such operation can do better. When $\eta_F=1$, the logical qubit can be recovered perfectly. When $\eta_F=0$, the two basis labels can still be preserved, but their coherence is lost from the accessible region; the optimal entanglement fidelity is then $1/2$.

We briefly describe the recovery operation. Write the polar decomposition as $T_F=\mathsf U_F\abs{T_F}$, where $\abs{T_F}=\sqrt{T_F^\dagger T_F}$ is positive semidefinite. Its trace is $\Tr\abs{T_F}=\norm{T_F}_1=\eta_F$. The partial isometry $\mathsf U_F$ can be extended to a unitary on $F$ in the finite-dimensional space considered here. We apply the controlled unitary
\begin{equation}
 U_{SF}=\ket0\bra0_S\otimes\id_F
        +\ket1\bra1_S\otimes \mathsf U_F.
 \label{eq:polarunitary}
\end{equation}
and then trace out $F$. The diagonal logical entries remain unchanged, while Eq.~\eqref{eq:achievability} shows that the off-diagonal entries acquire the factor $\Tr(T_F\mathsf U_F^\dagger)=\eta_F$. The resulting channel is
\begin{equation}
 \begin{pmatrix}
 \rho_{00}&\rho_{01}\\
 \rho_{10}&\rho_{11}
 \end{pmatrix}
 \longmapsto
 \begin{pmatrix}
 \rho_{00}&\eta_F\rho_{01}\\
 \eta_F\rho_{10}&\rho_{11}
 \end{pmatrix},
 \label{eq:recoveredchannel}
\end{equation}
where $\rho_{ij}$ denote the entries of the input logical density matrix. The operation depends on the known encoding, not on the unknown input state. Evaluating the entanglement fidelity using Eq.~\eqref{eq:entanglement-fidelity} reproduces Eq.~\eqref{eq:optimalrecovery}.

The same operator $T_F$ also determines the phase visibility $V(F)$ defined in Eq.~\eqref{eq:visibility}. From the off-diagonal block of $\rho_{SF}$ in Eq.~\eqref{eq:visibilityblock}, we find
\begin{equation}
 V(F)=2\sqrt{pq}\,\eta_F
     =2\sqrt{pq}\,[2F_Q(F)-1].
 \label{eq:visibilityrecovery}
\end{equation}
Thus, both phase visibility and optimal recovery fidelity depend on $\eta_F$. The visibility also contains the factor $2\sqrt{pq}$ from the particular prepared state. In contrast, $F_Q(F)$ tests the encoding with the maximally mixed logical input $\id_L/2$, so it does not depend on the physical branch probabilities $p$ and $q$.

For the three-qubit model, the operators in Eq.~\eqref{eq:threeTF} of Appendix~\ref{app:recovery} give
\begin{align}
 \eta_{\varnothing}
 &=\abs{c_\theta},
 \label{eq:etaempty}\\
 \eta_{E_1}
 &=\sqrt{1-s_\theta^2s_\chi^2},
 \label{eq:eta1}\\
 \eta_{E_2}
 &=\abs{c_\theta},
 \label{eq:eta2}\\
 \eta_{E_1E_2}
 &=1.
 \label{eq:eta12}
\end{align}
In particular, $\eta_{\varnothing}=\eta_{E_2}$. The reason is that tracing out $E_1$ removes all terms with $E_1=1$ from $T_{E_2}$, leaving $T_{E_2}=c_\theta\ket0\bra0$. Consequently, adding $E_2$ to $S$ does not improve recovery, even when $E_2$ can distinguish the system branches. By contrast, $\eta_{E_1E_2}=1$: with access to both environment qubits, the full unitary encoding can be reversed and the logical state recovered perfectly.

\subsection{Information in the inaccessible environment
limits quantum recovery}

We now examine how information about the system branches remaining in the inaccessible environment $\bar F$ limits recovery from $S$ and $F$. The relevant quantity is the fidelity between the two conditional states on $\bar F$.

For density matrices $\rho$ and $\sigma$, we use the root fidelity $f(\rho,\sigma)=\norm{\sqrt{\rho}\sqrt{\sigma}}_1$. It equals one for identical states and zero for states with orthogonal supports. Uhlmann's theorem relates this fidelity to the maximum overlap of purifications. Applying it to the two environment branches, as shown in Eq.~\eqref{eq:uhlmannproof}, gives~\cite{Uhlmann1976}
\begin{equation}
 \eta_F=f(\rho_{\bar F}^{(0)},\rho_{\bar F}^{(1)}).
 \label{eq:uhlmann}
\end{equation}
We use the root fidelity here; some conventions call its square the fidelity.

Combining Eq.~\eqref{eq:uhlmann} with the trace-distance--fidelity inequalities~\cite{Fuchs1999} and the definition of $D_Z(\bar F)$ in Eq.~\eqref{eq:record-distance}, we obtain
\begin{equation}
 1-\eta_F\leq D_Z(\bar F)
 \leq\sqrt{1-\eta_F^2}.
 \label{eq:fvg}
\end{equation}
Using $\eta_F=2F_Q(F)-1$ from Eq.~\eqref{eq:optimalrecovery}, the upper bound becomes
\begin{equation}
 D_Z(\bar F)^2+[2F_Q(F)-1]^2\leq1.
 \label{eq:generaltradeoff}
\end{equation}
The two terms refer to complementary regions: $D_Z(\bar F)$ measures how well the two system branches can be distinguished using the inaccessible environment, while $F_Q(F)$ measures recovery from $S$ and the accessible fragment. If $D_Z(\bar F)=1$, then $\eta_F=0$ and $F_Q(F)=1/2$. If $D_Z(\bar F)=0$, then $\eta_F=1$ and recovery is perfect. Thus, information about the system branch label remaining outside the accessible region limits recovery, although the full state still evolves unitarily.

Equation~\eqref{eq:generaltradeoff} is the standard trace-distance--fidelity bound expressed in recovery variables. It is also related to complementarity between interference and path information~\cite{Englert1996}. To illustrate it, consider the fanout protocol in Eq.~\eqref{eq:fanout-protocol}, where $\theta=\pi$ and $0\leq\chi\leq\pi$. Eqs.~\eqref{eq:dz1}, \eqref{eq:dz2}, and \eqref{eq:optimalrecovery}, together with the amplitudes in Eqs.~\eqref{eq:eta1}--\eqref{eq:eta12}, give
\begin{align}
 D_Z(E_1)&=1,\qquad
 D_Z(E_2)=\sin(\chi/2),
 \label{eq:fanrecords}\\
 F_Q(E_1)&=\frac{1+\abs{c_\chi}}2,\qquad
 F_Q(E_2)=\frac12,\qquad
 F_Q(E_1E_2)=1.
 \label{eq:fanrecovery}
\end{align}
Consider an observer who can access $S$ and $E_1$ but not $E_2$. At $\chi=0$, recovery from $SE_1$ is perfect. As $\chi$ increases, measurements on $E_2$ distinguish the system branches more easily, while $F_Q(E_1)$ decreases. Thus, the redistribution of the branch information to the inaccessible qubit limits recovery from $SE_1$. Throughout this process, the reduced state of $S$ remains unchanged.

Along the fanout protocol, the two conditional states of $E_2$ are $\ket0$ and $\ket{v_\chi}=c_\chi\ket0+s_\chi\ket1$. Their overlap is $c_\chi$, and Eqs.~\eqref{eq:eta1} and \eqref{eq:fanrecords} give
\begin{equation}
 D_Z(E_2)^2+[2F_Q(E_1)-1]^2=1.
 \label{eq:saturation}
\end{equation}
The upper bound of Eq.~\eqref{eq:fvg} is therefore saturated on this path. Using Eq.~\eqref{eq:visibilityrecovery}, the second term can also be written as $V(E_1)^2/(4pq)$.

For general $\theta$ and $\chi$, however, the corresponding expression is
\begin{equation}
 D_Z(E_2)^2+\eta_{E_1}^2
 =1-s_\theta^2c_\theta^2s_\chi^2.
 \label{eq:strictbound}
\end{equation}
Hence the bound need not be saturated away from the fanout path. Equality also holds at parameter values for which $s_\theta^2c_\theta^2s_\chi^2=0$.

\begin{table}[htbp]
\centering
\caption{The same system state with different
record and recovery properties.
Both columns have $\theta=\pi$, so
$\rho_S=q\ket0\bra0+p\ket1\bra1$.
Recovery always includes $S$ as well as
the listed fragment.}
\label{tab:fixedsystem}
\begin{tabular}{@{}lcc@{}}
\toprule
Quantity & Before fanout: $\chi=0$
         & Complete fanout: $\chi=\pi$\\
\midrule
System purity deficit $P_S$ & $2pq$ & $2pq$\\
Record $D_Z(E_1)$ & $1$ & $1$\\
Record $D_Z(E_2)$ & $0$ & $1$\\
Recovery $F_Q(E_1)$ & $1$ & $1/2$\\
Recovery $F_Q(E_1E_2)$ & $1$ & $1$\\
\bottomrule
\end{tabular}
\end{table}
Table~\ref{tab:fixedsystem} compares the beginning and the end of fanout. In both cases, the reduced state of $S$ is $q\ket0\bra0+p\ket1\bra1$, and $E_1$ distinguishes the branch labels perfectly. After complete fanout, $E_2$ also distinguishes the branches perfectly, but the recovery fidelity from $SE_1$ falls from one to $1/2$. Recovery is still perfect if both environment qubits are accessible. The example shows that the reduced system state alone does not specify where the branch information is available or which environment degrees of freedom are needed for quantum recovery.

\subsection{Branch distinguishability and quantum recovery
at complete fanout}

At complete fanout, each environment qubit carries the classical label. We compare the smallest fragments that can read that label with the fragments needed to recover the full quantum state.

At $\theta=\chi=\pi$, either environment qubit alone identifies the system label perfectly, but recovery of an unknown qubit requires access to both, in addition to $S$. An \emph{access structure} lists the sets sufficient to perform a specified task~\cite{Cleve1999,Gottesman2000}. A set is called authorized when it performs the task perfectly, and minimal when none of its proper subsets does so. Let $\mathcal H_Z^{\min}$ list the minimal environment sets with $D_Z=1$, and let $\mathcal H_Q^{\min}$ list those with $F_Q=1$. Then
\begin{equation}
 \mathcal H_Z^{\min}=\{\{E_1\},\{E_2\}\},\qquad
 \mathcal H_Q^{\min}=\{\{E_1,E_2\}\}.
 \label{eq:minimalsets}
\end{equation}
The symbols $\mathcal H_Z^{\min}$ and $\mathcal H_Q^{\min}$ are collections of subsets, not Hilbert spaces or Hamiltonians. Larger sets are also authorized because unused qubits can be discarded. The system $S$ is understood to be available for the recovery task only. Thus, these collections compare two tasks rather than defining a single quantum secret-sharing problem.

The pure state at this endpoint is a generalized Greenberger--Horne--Zeilinger (GHZ) state, a coherent superposition of all qubits having label zero and all having label one. It and its dephased counterpart are
\begin{align}
 \ket{\Psi_{\rm GHZ}}&=\sqrt q\ket{000}+e^{i\phi}\sqrt p\ket{111},\label{eq:ghz}\\
 \rho_{\rm mix}&=q\ket{000}\bra{000}+p\ket{111}\bra{111}.
 \label{eq:ghzmix}
\end{align}
Tracing out any one or more qubits removes the cross term $\ket{000}\bra{111}$. Thus every proper marginal, meaning a reduced state of fewer than all three qubits, is the same for these two global states~\cite{Walck2008}. This equality is exact, not an approximation from small signals. Nevertheless, the full three-qubit correlations depend on the phase. With $\langle O\rangle=\Tr(\ket\Psi\bra\Psi O)$, we obtain
\begin{align}
 \langle X_SX_{E_1}X_{E_2}\rangle&=2\sqrt{pq}\cos\phi,\label{eq:ghzx}\\
 \langle Y_SX_{E_1}X_{E_2}\rangle&=2\sqrt{pq}\sin\phi,\label{eq:ghzy}\\
 \langle X_SX_{E_1}X_{E_2}\rangle^2
 +\langle Y_SX_{E_1}X_{E_2}\rangle^2&=4pq=2P_S.
 \label{eq:ghzwitness}
\end{align}
The corresponding correlators vanish for $\rho_{\rm mix}$. To measure them, one can measure the indicated local Pauli operators on repeated preparations, multiply the three outcomes within each run, and average. Two settings, with $X_S$ or $Y_S$, give the two phase components. No entangling measurement is required, but the outcomes from all three sites must be combined. Without that joint information, the pure state and its mixture are indistinguishable.

Measuring these correlations diagnoses global coherence; it is not by itself a procedure that returns an unknown input qubit. Full coherent recovery is instead the channel task defined in Eq.~\eqref{eq:fqdef}. At this endpoint both tasks require access beyond every proper subsystem, while reading the classical label requires only one environment qubit.

\section{Entanglement and phase information during fanout}\label{sec:spectrum}
\subsection{Pairwise and tripartite entanglement during fanout}
The system purity stays fixed during fanout, but the form of entanglement changes. We calculate concurrence and three-tangle to see how pairwise entanglement is converted into GHZ-type tripartite entanglement.

We next ask whether the redistribution of recovery access can be related to entanglement measures. For two qubits $A,B$, write $C_{AB}=C(\rho_{AB})$ for the \emph{concurrence}. This is a normalized measure of pairwise entanglement: it is zero for separable two-qubit states and one for a maximally entangled pure pair. A separable state is a mixture of product states of the two qubits. The concurrence formula is given in Eq.~\eqref{eq:concurrencedef}~\cite{Wootters1998}. For the reduced states obtained from Eq.~\eqref{eq:sequential-final-state},
\begin{align}
 C_{SE_1}&=2\sqrt{pq}\abs{s_\theta c_\chi},\label{eq:cs1}\\
 C_{SE_2}&=0,\label{eq:cs2}\\
 C_{E_1E_2}&=2p\abs{s_\theta c_\theta s_\chi}.\label{eq:c12}
\end{align}
In particular, $C_{SE_2}=0$ at all pulse areas, as confirmed by the separable decomposition in Eq.~\eqref{eq:se2separable}. This is consistent with the distinction between pairwise entanglement and entanglement across $E_2|SE_1$. The concurrence $C_{E_1E_2}$ here is computed without conditioning on $S$; it is not the conditional purity in Eq.~\eqref{eq:conditionalpurity}.

For a pure three-qubit state, the \emph{three-tangle} is defined by the Coffman--Kundu--Wootters (CKW) relation in Eq.~\eqref{eq:tangledef}~\cite{CKW2000}. The determinant is denoted by $\det$. The subscript three labels this entanglement quantity; $\tau_3$ is not the pulse parameter $\tau$ or elapsed time. In our model,
\begin{equation}
 \tau_3=4pq\,s_\theta^2s_\chi^2.\label{eq:tangle}
\end{equation}
For this state family, $\tau_3$ describes the GHZ-type contribution not accounted for by the two pairwise concurrences involving $S$. It does not detect every possible form of three-qubit entanglement in general states. Using $1-\Tr\rho_S^2=2\det\rho_S$ and $C_{SE_2}=0$, the CKW relation becomes
\begin{equation}
 2P_S=C_{SE_1}^2+\tau_3.\label{eq:ckw}
\end{equation}
Along $\theta=\pi$, increasing $\chi$ changes
\begin{equation}
 C_{SE_1}^2=4pq\,c_\chi^2,\qquad
 \tau_3=4pq\,s_\chi^2,\label{eq:conversion}
\end{equation}
while $P_S$ remains fixed. At $\chi=0$, the system is entangled with $E_1$ as a pair and $E_2$ is uncorrelated. At $\chi=\pi$, the pairwise contribution vanishes and the same total system--environment entanglement appears in the GHZ-type term. Thus the unchanged mixedness of $S$ conceals a change in how its entanglement is shared.

\subsection{Phase-access increments and Pauli correlations}

We enlarge the accessible region one qubit at a time and ask how much extra phase becomes recoverable at each step. The increments are defined from recovery. Pauli strings are then used only as measurable witnesses of those increments.

Now fix a practical access procedure: the observer first controls $S$, then acquires $E_1$, and finally acquires $E_2$. The corresponding environment fragments are
\begin{equation}
 F_0=\varnothing,\quad F_1=E_1,\quad F_2=E_1E_2,
 \label{eq:forwardfiltration}
\end{equation}
For $k=0,1,2$, define the cumulative squared coherent recovery amplitude
\begin{equation}
 R_k=\eta_{F_k}^2=[2F_Q(F_k)-1]^2
                   =\frac{V(F_k)^2}{4pq}.\label{eq:rk}
\end{equation}
The scalar $R_k$ is distinct from the recovery map $\mathcal R$. It lies between zero and one and cannot decrease when more qubits become accessible: the observer can always ignore additional qubits. Since full access reverses the encoding, $R_2=1$. Separate the value available initially from the two gains,
\begin{equation}
 w_1=R_0,\qquad w_2=R_1-R_0,\qquad w_3=R_2-R_1
 \label{eq:wdef}
\end{equation}
so that $w_1,w_2,w_3$ are nonnegative and sum to one. The first entry is the squared coherence available from $S$ alone; the second is the extra amount enabled by adding $E_1$; the third is the further amount enabled by adding $E_2$. For this model,
\begin{equation}
 (w_1,w_2,w_3)
 =(c_\theta^2,s_\theta^2c_\chi^2,s_\theta^2s_\chi^2).
 \label{eq:wthree}
\end{equation}
The square is a choice of normalization, not an additional recovery theorem. It is useful here because the three increments equal the probabilities of $\ket{00}$, $\ket{10}$, and $\ket{11}$ in the conditional state $\ket{r_1}$. Another increasing function of $F_Q$ would give different increments. We refer to this ordered list as the \emph{phase-access spectrum}; ``spectrum'' here means a list of nonnegative weights, not an energy or frequency spectrum. Its index counts steps in the chosen access procedure. It also differs from multiple-quantum-coherence spectra, which classify density-matrix coherences by changes of a chosen collective observable~\cite{Gartner2018}.

To relate these normalized weights to measurable correlations, define $\Xi_r=4pq\,w_r$ for $r=1,2,3$. This restores the physical branch weights removed in $R_k$.

The $X$ operators below have a different purpose from the local $X_{E_2}$ readout in Section~\ref{sec:assumptions}. The local operator $X_{E_2}$ distinguishes two reduced states of one environment qubit. The joint products below connect the two global system branches and reveal their relative phase. They are witnesses of the recovery increments. They do not generate the dynamics and they do not define a general entanglement measure. For this circuit, the branch matrix elements in Eqs.~\eqref{eq:pauligeneral} and \eqref{eq:paulimatrixelements} give
\begin{align}
 \Xi_1&=\langle X_S\rangle^2+\langle Y_S\rangle^2,\label{eq:xi1}\\
 \Xi_2&=\langle X_SX_{E_1}\rangle^2+\langle Y_SX_{E_1}\rangle^2,\label{eq:xi2}\\
 \Xi_3&=\langle X_SX_{E_1}X_{E_2}\rangle^2
        +\langle Y_SX_{E_1}X_{E_2}\rangle^2.
 \label{eq:xi3}
\end{align}
These equalities provide a way to infer the increments from measured correlations under the assumed state model. The word witness here means an observable signal of the specified quantity, not a universal entanglement test for arbitrary states. The increments are defined by the change in recovery in Eq.~\eqref{eq:wdef}, rather than by counting Pauli factors. The spatial support of a Pauli product is the set of sites on which it is not the identity; adding diagonal factors can enlarge this support without providing new recoverable coherence. Thus operator support alone would give an ambiguous definition.

For the forward order in Eq.~\eqref{eq:forwardfiltration},
\begin{equation}
 \Xi_2=C_{SE_1}^2,\qquad \Xi_3=\tau_3,
 \qquad \Xi_2+\Xi_3=2P_S.
 \label{eq:xis}
\end{equation}
For the specified state and forward access order, these equations give an operational meaning to the two terms in the CKW relation: they are the gains in squared phase visibility enabled by the next available qubit. The equality depends on this state family, the quadratic normalization, and the access order. It is not a general identification of recovery gains with intrinsic entanglement measures.

\subsection{Dependence on the access order}
We repeat the recovery decomposition with the opposite order of access. This check shows that the increments describe an operational procedure and are not intrinsic multipartite entanglement measures.

To see the role of the access order, consider the same state but reverse the order in which the environment qubits are acquired,
\begin{equation}
 F'_0=\varnothing,\quad F'_1=E_2,\quad F'_2=E_1E_2.
 \label{eq:reversefiltration}
\end{equation}
Define $R'_k=\eta_{F'_k}^2$ and the primed increments by the same differences as in Eq.~\eqref{eq:wdef}. Equations~\eqref{eq:etaempty}, \eqref{eq:eta2}, and \eqref{eq:eta12} then give
\begin{equation}
 (w'_1,w'_2,w'_3)=(c_\theta^2,0,s_\theta^2).
 \label{eq:reverseweights}
\end{equation}
The middle increment is zero because $E_2$ adds no recovery power to $S$. The last increment then includes everything supplied by $E_1$, so in general $\Xi'_3\equiv4pq\,w'_3\ne\tau_3$. The prime labels a different access order for the \emph{same} physical state, not different dynamics. For example, at $\theta=\pi$ and $\chi=0$, the forward weights are $(0,1,0)$ and the reverse weights are $(0,0,1)$, although $\tau_3=0$. A nonzero final increment therefore need not represent three-party entanglement: in the reverse order, the essential qubit $E_1$ is simply acquired last.

The forward order is natural when an observer starts from $S$ and gradually enlarges a connected accessible region along the chain. However, this ordering is not determined by the abstract quantum state. A different laboratory geometry or access restriction should be described by a different sequence of fragments. Thus, the access order is part of the operational setup.

\section{Generalization to a chain of $N$ environment qubits}
\label{sec:chain}
\subsection{Conditional states of the environment chain}
We first solve the conditional branch once for an arbitrary chain. The prefix amplitudes obtained here will determine every later record and recovery formula.

The chain extension shows how the distinction between entanglement and local records depends on distance along the specified sequence of interactions. There are $N$ environment qubits, with $N\geq1$, in addition to the system, so the total is $N+1$. We use $E_0\equiv S$ to write every gate in the same form. The system starts in $\sqrt q\ket0+e^{i\phi}\sqrt p\ket1$, and every $E_j$, $j\geq1$, starts in $\ket0$. Apply the gates in the order $1,2,\ldots,N$:
\begin{equation}
 U_j(\theta_j)=\ket0\bra0_{E_{j-1}}\otimes\id_{E_j}
       +\ket1\bra1_{E_{j-1}}\otimes R_y^{E_j}(\theta_j).
 \label{eq:chaingates}
\end{equation}
The total unitary is $U_N(\theta_N)\cdots U_1(\theta_1)$, with $U_1$ acting first. The branch $z=0$ remains $\ket{r_0^{(N)}}=\ket{0^N}$, where $\ket{0^N}=\ket0^{\otimes N}$. The superscript $(N)$ denotes the number of environment qubits. Define $c_j=\cos(\theta_j/2)$ and $s_j=\sin(\theta_j/2)$. The other branch is
\begin{equation}
 \ket{r_1^{(N)}}=\sum_{j=0}^{N}A_j\ket{1^j0^{N-j}},
 \label{eq:prefixstate}
\end{equation}
where $\ket{1^j0^{N-j}}$ means that the first $j$ environment qubits are one and the remaining $N-j$ are zero. Such a string is called a \emph{prefix of ones}; for example, at $N=4$ the allowed strings are $0000$, $1000$, $1100$, $1110$, and $1111$. The coefficient $A_j$ is the amplitude of the prefix with $j$ ones:
\begin{align}
 A_0&=c_1,\label{eq:a0}\\
 A_j&=\left(\prod_{i=1}^{j}s_i\right)c_{j+1},
       \quad1\leq j<N,\label{eq:ak}\\
 A_N&=\prod_{i=1}^{N}s_i.\label{eq:an}
\end{align}
For $j<N$, the product of sines describes activation of the first $j$ targets, and $c_{j+1}$ is the amplitude for the next target to remain zero. For $j=N$, no further target remains, so there is no final cosine. In each basis component, a zero leaves the next control inactive and prevents further propagation. The state is still a coherent superposition of all these possible prefix lengths; no intermediate measurement is assumed.

Normalization, $\sum_{j=0}^N|A_j|^2=1$, follows by repeatedly combining $c_j^2+s_j^2=1$. The telescoping derivation is given in Eq.~\eqref{eq:telescoping}. The branch overlap is $\braket{r_0^{(N)}}{r_1^{(N)}}=A_0=c_1$, so the system's reduced state depends only on the first pulse. All later pulses can redistribute records and recovery access without changing $\rho_S$.

\subsection{Entanglement and local record onset orders}\label{sec:chainonset}
For one site $E_k$, we separate the conditional population $B_k$ from the conditional coherence $u_k$. The $Z$ readout sees $B_k$, while the $X$ readout sees $u_k$. This is why the two detectors have different weak-signal powers.

To compute what a single site $E_k$, $1\leq k\leq N$, records, define its conditional excitation probability $B_k$ and conditional local coherence $u_k$ by
\begin{equation}
 B_k=\sum_{j=k}^{N}\abs{A_j}^2
       =\prod_{i=1}^{k}s_i^2,\qquad
 u_k=A_{k-1}A_k^*.\label{eq:bu}
\end{equation}
The sum for $B_k$ includes all prefixes long enough to reach site $k$. The product form follows from the telescoping identity in Eq.~\eqref{eq:telescoping}, applied from site $k$ onward. The complex conjugate is denoted by $*$. With these definitions, the conditional single-qubit states are
\begin{equation}
 \rho_{E_k}^{(1)}=\begin{pmatrix}1-B_k&u_k\\u_k^*&B_k\end{pmatrix},
 \qquad\rho_{E_k}^{(0)}=\ket0\bra0.
 \label{eq:chainrho}
\end{equation}
The off-diagonal entry survives a partial trace only when the two interfering basis strings agree at every other site. Among prefixes, this happens only for lengths $k-1$ and $k$, which explains $u_k=A_{k-1}A_k^*$. The difference $\rho_{E_k}^{(0)}-\rho_{E_k}^{(1)}$ has eigenvalues $\pm\sqrt{B_k^2+|u_k|^2}$, giving
\begin{equation}
 D_Z(E_k)=\sqrt{B_k^2+\abs{A_{k-1}A_k}^2}.
 \label{eq:chainrecord}
\end{equation}
Without conditioning on the system, $\rho_{E_k}=q\ket0\bra0+p\rho_{E_k}^{(1)}$. Its determinant gives the purity deficit
\begin{equation}
 P_{E_k}=2pB_k(1-pB_k)-2p^2\abs{u_k}^2.
 \label{eq:chainpurity}
\end{equation}
Equations~\eqref{eq:chainrecord} and \eqref{eq:chainpurity} are exact at any pulse areas. The former measures the distinction between two conditional states; the latter measures mixedness of their weighted average. Their different dependence on $B_k$ and $u_k$ is the algebraic reason their leading powers need not coincide. The partial-trace, eigenvalue, and determinant calculations underlying Eqs.~\eqref{eq:chainrho}--\eqref{eq:chainpurity} are detailed in Appendix~\ref{app:chain}.

\begin{proposition}[Entanglement and record orders]\label{prop:orders}
Assume the initial product preparation above, fixed $0<p<1$, the ordered gates in Eq.~\eqref{eq:chaingates}, and real analytic pulse areas satisfying
\begin{equation}
 \begin{gathered}
 \theta_j(\tau)=2g_j\tau^{\ell_j}+O(\tau^{\ell_j+1}),\qquad
 g_j\ne0,\qquad\ell_j\in\mathbb N_{>0},\qquad\tau\to0^+.
 \end{gathered}
 \label{eq:nonuniformpulses}
\end{equation}
Here $\mathbb N_{>0}=\{1,2,\ldots\}$, $\ell_j$ is the first nonzero power of the $j$th pulse, and $g_j$ is its real leading coefficient. If $\tau$ has units of time, $g_j$ has units of time to the power $-\ell_j$; for nonlinear pulse families it is not necessarily a constant Hamiltonian coupling. Define $M_0=0$ and $M_k=\sum_{j=1}^k\ell_j$, the accumulated pulse order up to site $k$. Define also $G_0=1$ and $G_k=\prod_{j=1}^k g_j$, the corresponding product of coefficients. With unrestricted local record measurements,
\begin{equation}
m_\star(E_k)=M_k,\qquad
 n_Z(E_k)=M_{k-1}+M_k.
 \label{eq:generalorderlaw}
\end{equation}
The leading coefficient is also obtained explicitly as
\begin{equation}
 D_Z(E_k)=\abs{G_{k-1}G_k}\,
          \tau^{M_{k-1}+M_k}[1+O(\tau)].
 \label{eq:generalrecordcoefficient}
\end{equation}
\end{proposition}

\begin{proof}
Each sine contributes its leading factor $g_j\tau^{\ell_j}$, while each cosine tends to one. Hence Eqs.~\eqref{eq:a0}--\eqref{eq:an} give, with the conventions for $k=0$ stated above,
\begin{equation}
 A_k=G_k\tau^{M_k}[1+O(\tau)],\qquad
 B_k=G_k^2\tau^{2M_k}[1+O(\tau)].
 \label{eq:prefixasymptotic}
\end{equation}
In particular, $u_k=G_{k-1}G_k\tau^{M_{k-1}+M_k}[1+O(\tau)]$. The two terms under the square root in Eq.~\eqref{eq:chainrecord} have orders $4M_k$ and $2(M_{k-1}+M_k)$. Since $M_k-M_{k-1}=\ell_k>0$, the local coherence term has the lower power and a nonzero coefficient. Taking the square root proves Eq.~\eqref{eq:generalrecordcoefficient}.

For $k=1$, the two leading terms in Eq.~\eqref{eq:chainpurity} partially cancel, leaving
\begin{equation}
 P_{E_1}=2pqg_1^2\tau^{2\ell_1}[1+O(\tau)].
 \label{eq:generalpurityfirst}
\end{equation}
For $k\geq2$, $M_{k-1}>0$ makes both subtractions higher order than $B_k$, so
\begin{equation}
 P_{E_k}=2pG_k^2\tau^{2M_k}[1+O(\tau)].
 \label{eq:generalpuritylater}
\end{equation}
A qubit has at most two nonzero Schmidt amplitudes; this number is its Schmidt rank. The positive leading coefficient of $P_{E_k}$ therefore fixes the smaller amplitude through Eq.~\eqref{eq:purity-scaling}, giving $m_\star(E_k)=M_k$. Thus the entanglement order follows from the reduced-state determinant, rather than from an assumption that every nonzero excitation amplitude is entanglement.
\end{proof}

For exact linear pulses $\theta_j=2g_j\tau$, the trigonometric parity improves the remainder to two further powers:
\begin{equation}
 D_Z(E_k)=
 \left(\prod_{j=1}^{k-1}g_j^2\right)\abs{g_k}\,
 \tau^{2k-1}+O(\tau^{2k+1}).
 \label{eq:linearrecordlaw}
\end{equation}
For linear pulses, reaching site $k$ requires an amplitude of order $\tau^k$. A locally readable off-diagonal element must multiply that amplitude by the amplitude of the neighboring prefix of length $k-1$, which is of order $\tau^{k-1}$. This produces the record power $2k-1$. The gap between record order and Schmidt order is therefore $k-1$, or $M_{k-1}$ for unequal pulse powers. It counts the extra suppression from the earlier activated controls in this circuit, not a general propagation time fixed by distance alone.

The physical origin of the separation is the same at every site. The conditional variance identity becomes
\begin{equation}
 P_{E_k}=p\,\mathcal P(\rho_{E_k}^{(1)})+2pqD_Z(E_k)^2.
 \label{eq:chainvariance}
\end{equation}
For $k\geq2$, the conditional-environment contribution starts at order $2M_k$, while the term associated directly with branch distinction starts at order $2(M_{k-1}+M_k)$. This is precisely the chain generalization of the mechanism found at $E_2$.

It is also possible to write the optimal local measurement explicitly. For the real amplitudes considered here, Eq.~\eqref{eq:chainrho} gives
\begin{equation}
 \Delta_k\equiv\rho_{E_k}^{(0)}-\rho_{E_k}^{(1)}
              =B_k Z_{E_k}-u_k X_{E_k}.
 \label{eq:localrecordoperator}
\end{equation}
When $D_Z(E_k)>0$, define
\begin{equation}
 O_k=\frac{B_k Z_{E_k}-u_k X_{E_k}}
               {\sqrt{B_k^2+u_k^2}},\qquad
 \Pi_\pm=\frac{\id\pm O_k}{2}.
 \label{eq:optimalrecordmeasurement}
\end{equation}
The operator $\Delta_k$ is the difference of conditional density matrices, not the dephasing map $\Delta_Z$. The normalized observable $O_k$ has eigenvalues $\pm1$, and $\Pi_\pm$ are the projectors onto its two eigenspaces. Measuring them and assigning outcomes $+$ and $-$ to labels zero and one gives success probability $(1+D_Z(E_k))/2$. This optimal two-state discrimination is called the Helstrom measurement. If $D_Z(E_k)=0$, the conditional states are identical and the best equal-prior success probability is $1/2$.

Although the exact optimal direction depends on the pulse areas, this dependence is unnecessary at leading order. The fixed transverse algebra $\cA_x(E_k)=\spanop\{\id,X_{E_k}\}$ gives
\begin{equation}
 D_Z^{\cA_x}(E_k)=\frac12\abs{\Tr(X_{E_k}\Delta_k)}
                =\abs{u_k}.
 \label{eq:xrestrictedchain}
\end{equation}
Under the assumptions of Proposition~\ref{prop:orders}, $B_k/\abs{u_k}\to0$. Therefore $D_Z^{\cA_x}(E_k)/D_Z(E_k)\to1$. A fixed $X$ measurement not only reproduces the onset exponent but also the leading coefficient,
\begin{equation}
 n_Z^{\cA_x}(E_k)=M_{k-1}+M_k.
 \label{eq:xrecordorder}
\end{equation}
At finite pulse areas, of course, the exact Helstrom direction in Eq.~\eqref{eq:optimalrecordmeasurement} need not coincide with the $X$ axis.

If the observer is instead restricted to the population algebra
$\cA_z(E_k)=\spanop\{\id,Z_{E_k}\}$, Eq.~\eqref{eq:chainrho} instead gives
\begin{equation}
 D_Z^{\cA_z}(E_k)=B_k,
 \qquad n_Z^{\cA_z}(E_k)=2M_k.
 \label{eq:restrictedchainlaw}
\end{equation}
For linear pulses the comparison is therefore
\begin{equation}
 \begin{gathered}
 m_\star(E_k)=k,\qquad n_Z^{\cA_z}(E_k)=2k,\qquad
 n_Z(E_k)=n_Z^{\cA_x}(E_k)=2k-1.
 \end{gathered}
 \label{eq:threeorders}
\end{equation}
The comparison uses one state and one dynamics. The difference between $m_\star$ and $n_Z$ reflects the different tasks of quantifying entanglement and reading a record. The difference between the $X$ and population record orders is instead due entirely to the allowed readout. A smaller Schmidt order should not be mistaken for a claim that every entanglement measurement detects a signal at that same power.

Proposition~\ref{prop:orders} applies to the specified ready states and ordered pulse family. Vanishing leading couplings, different initial states, or additional interactions can change the orders. If $p=0$ or $p=1$, the physical superposition of two system branches is absent and some entanglement orders change; the equal-prior distance between two hypothetical conditional preparations can still be defined separately. The ready-state counterexample in Eqs.~\eqref{eq:misalignedrecord} and \eqref{eq:misalignedpurity} shows why connectivity alone is not sufficient. The fixed-Hamiltonian result establishes the second-site separation in physical time, but is not a theorem for a simultaneously interacting chain of arbitrary length.

\subsection{Recovery from an accessible prefix}\label{sec:chainrecovery}
We calculate which prefix amplitudes remain coherent after the rest of the chain is discarded. Their total weight gives the optimal recovery from the accessible prefix.

For the recovery task, let the observer enlarge a connected accessible region starting from $S$. The environment part of this region is
\begin{equation}
 F_k=E_1\cdots E_k,\quad 0\leq k\leq N,
 \qquad F_0=\varnothing.
 \label{eq:chainprefix}
\end{equation}
Thus recovery at step $k$ uses $S,E_1,\ldots,E_k$, while the inaccessible qubits are $E_{k+1},\ldots,E_N$. The reference branch $\ket{0^N}$ has no ones. A term $\ket{1^j0^{N-j}}$ with $j>k$ differs from it on at least one lost qubit, so its off-diagonal cross term vanishes under the partial trace. Only terms with $j\leq k$ remain, giving
\begin{equation}
 T_{F_k}=\ket{0^k}\sum_{j=0}^k A_j^*\bra{1^j0^{k-j}},
 \label{eq:chainTF}
\end{equation}
For $k=0$, the empty tensor product is the scalar one, so $T_{F_0}=A_0^*$. For larger $k$, $T_{F_k}$ is an outer product of a unit vector and a generally unnormalized vector. It has at most one nonzero singular value, equal to that second vector's norm:
\begin{equation}
 \eta_{F_k}=\sqrt{\sum_{j=0}^k\abs{A_j}^2}.
 \label{eq:chaineta}
\end{equation}
If this norm vanishes, the operator is zero; the same formula still applies. Hence
\begin{equation}
 [2F_Q(F_k)-1]^2
       =\frac{V(F_k)^2}{4pq}
       =\sum_{j=0}^{k}\abs{A_j}^2.
 \label{eq:chainrecovery}
\end{equation}
The sum contains precisely the conditional components that leave every lost qubit in the reference ready state. They can therefore retain coherence with the zero branch after those lost qubits are discarded. This gives a direct interpretation of the cumulative recovery formula. Appendix~\ref{app:chain} supplies the partial-trace and singular-value steps leading to Eqs.~\eqref{eq:chainTF}--\eqref{eq:chainrecovery}. The associated increments in squared recoverable coherence are
\begin{equation}
 w_1=\abs{A_0}^2,\qquad
 w_{k+1}=\abs{A_k}^2\quad(1\leq k\leq N),
 \label{eq:chainweights}
\end{equation}
They satisfy $\sum_{r=1}^{N+1}w_r=1$. The index $r=1$ denotes access to $S$ alone; $r=k+1$ denotes the additional squared coherence enabled when $E_k$ is added to the already accessible prefix. Increasing this index changes the accessible region for a fixed encoded state, not elapsed time. These increments are neither site populations in the full state nor intrinsic multipartite entanglement orders.

For the pulse family in Proposition~\ref{prop:orders}, $w_{k+1}$ starts at power $2M_k=2m_\star(E_k)$. This differs from the local record power $M_{k-1}+M_k$ because the tasks use different subsystems: prefix recovery uses $S$ and the first $k$ environment qubits, whereas local record readout uses $E_k$ alone. The equality of one recovery exponent with a purity exponent in this model does not identify the two observables.

For the real amplitudes in Eq.~\eqref{eq:prefixstate}, the increments can also be read from the Pauli products
\begin{equation}
 X^{(k)}=X_S\prod_{j=1}^{k}X_{E_j},\qquad
 Y^{(k)}=Y_S\prod_{j=1}^{k}X_{E_j}.
 \label{eq:chainpaulis}
\end{equation}
Here $0\leq k\leq N$, and an empty product at $k=0$ is the identity, giving $X^{(0)}=X_S$ and $Y^{(0)}=Y_S$. These products connect the reference branch only to the prefix of length $k$. Their expectation values are therefore $2\sqrt{pq}A_k\cos\phi$ and $2\sqrt{pq}A_k\sin\phi$. With $\Xi_{k+1}=4pq\,w_{k+1}$,
\begin{equation}
 \langle X^{(k)}\rangle^2+\langle Y^{(k)}\rangle^2
 =4pq\abs{A_k}^2=\Xi_{k+1}.
 \label{eq:chainwitness}
\end{equation}
These correlation signals test the recovery increments for the stated model and access order. They do not define a general measure of $(k+1)$-party entanglement. At complete fanout, every $\theta_j=\pi$, so only $A_N$ survives and $w_{N+1}=1$. Every environment qubit then carries a perfect copy of the branch label, but coherence is visible only with access to the full joint correlation. The chain extends the three-qubit distinction between many local classical records and one globally recoverable quantum state. The Pauli-correlation derivation of Eq.~\eqref{eq:chainwitness} is included in Appendix~\ref{app:chain}.

\section{Summary}\label{sec:discussion}

This paper used one solvable qubit chain to answer three different questions.

First, we calculated when each site becomes entangled with the rest. The relevant quantity is the smaller Schmidt amplitude across a specified cut, obtained from the purity deficit. For linear weak pulses, the site $E_k$ enters at amplitude order $k$.

Second, we calculated when the same site contains a readable record of the system's $Z$-branch. The relevant quantity is the trace distance between the two conditional states of that site. The optimal local record at $E_k$ starts at order $2k-1$. A fixed local $X_{E_k}$ measurement reads this leading signal because it is stored in an off-diagonal matrix element. A population-only $Z_{E_k}$ measurement starts at order $2k$. These three powers answer different questions and should not be combined into one notion of ``information order.''

Third, we calculated whether an unknown logical qubit can be recovered from the system and an accessible fragment. The answer is controlled by the surviving branch operator $T_F$ and its trace norm. During environment-only fanout, the reduced state of $S$ remains fixed, but a second readable record grows outside a chosen accessible region and the recovery fidelity inside that region falls. Full recovery remains possible when all qubits are controlled. This is redistribution of information in a reversible global state, not fundamental loss of information.

%The two appearances of the Pauli $X$ operator should be kept separate. A single-site $X_{E_k}$ is a local record readout. A product such as $X_SX_{E_1}\cdots X_{E_k}$, together with its $Y_S$ quadrature, is a joint phase witness. The first reads a branch-dependent local coherence. The second shows where the relative phase is visible in a larger region. Neither operator creates the dynamics; the controlled rotations are generated by $Y$.

%The phase-access increments are defined from changes in optimal recovery as the accessible prefix grows. In the three-qubit state family, these increments happen to equal a squared concurrence and the three-tangle for the forward access order. Reversing the access order changes the increments. They are therefore operational quantities, not a new intrinsic decomposition of multipartite entanglement.

%The model is finite, coherent, and reversible. It does not describe irreversible decoherence by itself. Its role is to provide an exact benchmark: local mixedness, readable records, and recoverable coherence are related, but no single reduced-state scalar determines all three. A claim of effective classicality must specify the state, the subsystem split, the observable algebra, and the controls available to the observer.

\section*{Acknowledgments}
S. K. was supported by the Japan Society for the Promotion of Science (JSPS) KAKENHI Grant No. JP24K21548. J. S. was supported in part by JSPS KAKENHI Grants No. JP23K22491, No. JP24K21548, and No. JP25H02186.

\appendix
\section{The fixed-Hamiltonian expansion}\label{app:fixed}
We derive the short-time expressions used in Section~4.3. In the ordered environment basis $\ket{00},\ket{01},\ket{10},\ket{11}$, the conditional Hamiltonian in Eq.~\eqref{eq:conditionalh} is
\begin{equation}
 H^{(1)}=\begin{pmatrix}
 0&0&-ig&0\\
 0&0&0&-ig\\
 ig&0&0&-i\lambda\\
 0&ig&i\lambda&0
 \end{pmatrix}.\label{eq:hmat}
\end{equation}
Acting on the initial environment state gives
\begin{align}
 H^{(1)}\ket{00}&=ig\ket{10},\nonumber\\
 (H^{(1)})^2\ket{00}&=g^2\ket{00}-g\lambda\ket{11},\nonumber\\
 (H^{(1)})^3\ket{00}&=ig(g^2+\lambda^2)\ket{10}
                         +ig^2\lambda\ket{01}.
 \label{eq:hpowers}
\end{align}
Substitution into $\exp(-iH^{(1)}t)\ket{00}$ gives Eq.~\eqref{eq:fixed-hamiltonian-state}. In particular, the $\ket{11}$ and $\ket{01}$ amplitudes start as $g\lambda t^2/2$ and $-g^2\lambda t^3/6$, respectively. The latter amplitude is absent in the sequential-pulse protocol.

To calculate the state of $E_2$, write $b_{ab}(t)=\bra{ab}r_1^{\rm on}(t)\rangle$, where $a$ and $b$ label $E_1$ and $E_2$. The conditional probability of $E_2=1$ and its off-diagonal element are
\begin{align}
 B(t)&=\abs{b_{01}}^2+\abs{b_{11}}^2
       =\frac{g^2\lambda^2}{4}t^4+O(t^6),\label{eq:fixedB}\\
 u(t)&=b_{00}b_{01}^*+b_{10}b_{11}^*
       =\frac{g^2\lambda}{3}t^3+O(t^5).\label{eq:fixedu}
\end{align}
The two contributions to $u(t)$ are
\[
 b_{00}b_{01}^*=-\frac{g^2\lambda}{6}t^3+O(t^5),\qquad
 b_{10}b_{11}^*=\frac{g^2\lambda}{2}t^3+O(t^5).
\]
Both are needed to obtain the coefficient $1/3$. The amplitudes are real for the chosen initial state and Hamiltonian. Moreover, $H^{(1)}$ changes the excitation-number parity at each step. The amplitudes $b_{00}$ and $b_{11}$ are even in $t$, while $b_{01}$ and $b_{10}$ are odd. This gives the remainder powers in Eqs.~\eqref{eq:fixedB} and \eqref{eq:fixedu}.

Since the $S=0$ branch leaves $E_2$ in $\ket0$, the difference between its two conditional reduced states has the form of Eq.~\eqref{eq:tracelessmatrix}. Its trace distance is
\begin{equation}
 D_Z^{\rm on}(E_2;t)=\sqrt{B(t)^2+\abs{u(t)}^2}.
 \label{eq:fixedDproof}
\end{equation}
Equations~\eqref{eq:fixedB} and \eqref{eq:fixedu} then give Eq.~\eqref{eq:fixedrecord}. The unconditional state of $E_2$ has population $pB(t)$ and coherence $pu(t)$, so
\[
 \det\rho_{E_2}=pB(t)[1-pB(t)]-p^2\abs{u(t)}^2
 =\frac{pg^2\lambda^2}{4}t^4+O(t^6).
\]
Using $P_{E_2}=2\det\rho_{E_2}$ proves Eq.~\eqref{eq:fixedpe2}.

For completeness, the branch overlap obtained from Eq.~\eqref{eq:fixed-hamiltonian-state} is
\[
 \braket{r_0}{r_1^{\rm on}(t)}=b_{00}(t)
 =1-\frac{g^2t^2}{2}+O(t^4).
\]
The reduced-state determinant is therefore
\[
 \det\rho_S=pq\bigl(1-\abs{b_{00}(t)}^2\bigr)
 =pqg^2t^2+O(t^4),
\]
which gives Eq.~\eqref{eq:fixedps}. Similarly, the conditional state of $E_1$ has
\[
 (\rho_{E_1}^{(1)})_{11}=\abs{b_{10}}^2+\abs{b_{11}}^2
 =g^2t^2+O(t^4),\qquad
 (\rho_{E_1}^{(1)})_{01}=b_{00}b_{10}^*+b_{01}b_{11}^*
 =gt+O(t^3).
\]
Because $\rho_{E_1}=q\ket0\bra0+p\rho_{E_1}^{(1)}$, we find
\[
 \det\rho_{E_1}=pqg^2t^2+O(t^4).
\]
This proves Eq.~\eqref{eq:fixedpe1}. Thus all four short-time expressions in Eqs.~\eqref{eq:fixedps}--\eqref{eq:fixedrecord} follow from the same conditional state.

\section{The reduced states and the mixedness identity}\label{app:reduced}
We calculate the reduced states used in Section~4 and derive the identity that separates conditional mixedness from distinguishability of the two system branches.

Write $a_{ijk}=\bra{ijk}\Psi\rangle$, with $i,j,k\in\{0,1\}$ labeling $S,E_1,E_2$. The nonzero amplitudes of Eq.~\eqref{eq:sequential-final-state} are
\begin{align}
 a_{000}&=\sqrt q,&a_{100}&=e^{i\phi}\sqrt p\,c_\theta,\nonumber\\
 a_{110}&=e^{i\phi}\sqrt p\,s_\theta c_\chi,&
 a_{111}&=e^{i\phi}\sqrt p\,s_\theta s_\chi.
 \label{eq:amplitudes}
\end{align}
The matrix elements of a reduced state are obtained by summing over the discarded indices. For example,
\begin{equation}
 (\rho_S)_{ii'}=\sum_{j,k=0}^1a_{ijk}a_{i'jk}^*,\qquad
 (\rho_{E_2})_{kk'}=\sum_{i,j=0}^1a_{ijk}a_{ijk'}^*.
 \label{eq:partialtracecomponents}
\end{equation}
In particular, $(\rho_S)_{01}=\sqrt{pq}e^{-i\phi}c_\theta$, while $(\rho_{E_2})_{01}=p s_\theta^2c_\chi s_\chi$. Together with Eq.~\eqref{eq:rhos}, the remaining reduced states are
\begin{align}
 \rho_{E_1}&=\begin{pmatrix}
 q+pc_\theta^2&pc_\theta s_\theta c_\chi\\
 pc_\theta s_\theta c_\chi&ps_\theta^2
 \end{pmatrix},\label{eq:rhoe1}\\
 \rho_{E_2}&=\begin{pmatrix}
 1-ps_\theta^2s_\chi^2&ps_\theta^2c_\chi s_\chi\\
 ps_\theta^2c_\chi s_\chi&ps_\theta^2s_\chi^2
 \end{pmatrix}.\label{eq:rhoe2}
\end{align}
For a normalized qubit state, $1-\Tr\rho^2=2\det\rho$. Evaluation of the three determinants gives
\begin{align}
 \det\rho_S&=pq\,s_\theta^2,\nonumber\\
 \det\rho_{E_1}&=ps_\theta^2(q+pc_\theta^2s_\chi^2),\nonumber\\
 \det\rho_{E_2}&=ps_\theta^2s_\chi^2(1-ps_\theta^2),
 \label{eq:determinants}
\end{align}
which prove Eqs.~\eqref{eq:ps}--\eqref{eq:pe2}.

The conditional states are obtained by tracing the normalized environment branch $\ket{r_1}$ in Eq.~\eqref{eq:sequential-conditional-states}, without its probability factor $p$. They are
\begin{align}
 \rho_{E_1}^{(1)}&=\begin{pmatrix}
 c_\theta^2&c_\theta s_\theta c_\chi\\
 c_\theta s_\theta c_\chi&s_\theta^2
 \end{pmatrix},\nonumber\\
 \rho_{E_2}^{(1)}&=\begin{pmatrix}
 1-s_\theta^2s_\chi^2&s_\theta^2c_\chi s_\chi\\
 s_\theta^2c_\chi s_\chi&s_\theta^2s_\chi^2
 \end{pmatrix}.\label{eq:conditionalmatrices}
\end{align}
For $S=0$, both environment qubits have conditional state $\ket0\bra0$. The difference between this state and a conditional qubit state can be written as
\begin{equation}
 \begin{pmatrix}B&-u\\-u^*&-B\end{pmatrix},
 \label{eq:tracelessmatrix}
\end{equation}
where $B$ is real. Its eigenvalues are $\pm\sqrt{B^2+\abs{u}^2}$, giving the trace distance $\sqrt{B^2+\abs{u}^2}$. For $E_2$, $B=s_\theta^2s_\chi^2$ and $u=s_\theta^2c_\chi s_\chi$, so $D_Z(E_2)=s_\theta^2\abs{s_\chi}$, as in Eq.~\eqref{eq:dz2}. The same calculation using $\rho_{E_1}^{(1)}$ gives Eq.~\eqref{eq:dz1}. For the full environment, the two conditional states are pure and their trace distance is $\sqrt{1-\abs{\braket{r_0}{r_1}}^2}=\abs{s_\theta}$.

We next separate mixedness within a system branch from the effect of mixing different branches. For normalized density matrices $\rho_0,\rho_1$ and $p+q=1$,
\begin{equation}
 \Tr(q\rho_0+p\rho_1)^2
 =q^2\Tr\rho_0^2+p^2\Tr\rho_1^2
       +2pq\Tr(\rho_0\rho_1).
 \label{eq:purityexpansion}
\end{equation}
Using $\cP(\rho)=1-\Tr\rho^2$ and rearranging gives
\begin{equation}
 \begin{split}
 \cP(q\rho_0+p\rho_1)
 ={}q\cP(\rho_0)+p\cP(\rho_1)
 +pq\Tr[(\rho_0-\rho_1)^2].
 \end{split}\label{eq:variancegeneral}
\end{equation}
This equality holds in any finite dimension. For qubits, the traceless Hermitian difference $\rho_0-\rho_1$ has eigenvalues $\pm D$, where $D=\frac12\norm{\rho_0-\rho_1}_1$. Hence $\Tr[(\rho_0-\rho_1)^2]=2D^2$. In the present model, $\rho_0=\ket0\bra0$ and $\cP(\rho_0)=0$, yielding Eq.~\eqref{eq:conditionalvariance}. Taking the determinant of $\rho_{E_2}^{(1)}$ gives $\det\rho_{E_2}^{(1)}=s_\theta^2c_\theta^2s_\chi^2$, which proves Eq.~\eqref{eq:conditionalpurity}. The replacement of $\Tr[(\rho_0-\rho_1)^2]$ by $2D^2$ is specific to qubits.

Finally, consider the changed preparation of $E_1$ in Section~\ref{sec:assumptions}. The $S=0$ branch leaves $E_1$ in $\ket+$, while the $S=1$ branch rotates it to $R_y(\theta)\ket+=[(c_\theta-s_\theta)\ket0+(c_\theta+s_\theta)\ket1]/\sqrt2$. The probabilities that $E_1=1$ before the second gate are
\begin{equation}
 a_0=\frac12,\qquad a_1=\frac{1+\sin\theta}{2}.
 \label{eq:misalignedprobabilities}
\end{equation}
Writing $\ket{v_\chi}=c_\chi\ket0+s_\chi\ket1$, the corresponding conditional states of $E_2$ are
\begin{equation}
 \rho_{E_2}^{(z)}=(1-a_z)\ket0\bra0+a_z\ket{v_\chi}\bra{v_\chi}.
 \label{eq:misalignedconditional}
\end{equation}
Since the distance between $\ket0\bra0$ and $\ket{v_\chi}\bra{v_\chi}$ is $\abs{s_\chi}$, the difference $\abs{a_1-a_0}$ gives Eq.~\eqref{eq:misalignedrecord}. The unconditional state has $a=qa_0+pa_1=(1+p\sin\theta)/2$ and determinant $a(1-a)s_\chi^2$, proving Eq.~\eqref{eq:misalignedpurity}.

\section{The optimal recovery formula}\label{app:recovery}
We derive Eq.~\eqref{eq:optimalrecovery} by constructing a recovery operation that reaches the stated fidelity and proving that no deterministic recovery operation can exceed it.

Let $R$ and $L$ be the reference and logical qubits introduced in Eq.~\eqref{eq:entfidelitydefinition}, and take
\begin{equation}
 \ket\Phi_{RL}=(\ket{00}+\ket{11})/\sqrt2.
 \label{eq:bellinput}
\end{equation}
The encoding in Eq.~\eqref{eq:encoding} induces the channel
\begin{equation}
 \mathcal N_F(\ket i\bra j)=\ket i\bra j_S\otimes T_{ij}^{(F)},
 \label{eq:channelblocks}
\end{equation}
where $T_{ij}^{(F)}=\Tr_{\bar F}\ket{r_i}\bra{r_j}$, so $T_{00}^{(F)}=\rho_F^{(0)}$, $T_{11}^{(F)}=\rho_F^{(1)}$, and $T_{01}^{(F)}=T_F$. Expanding the entanglement fidelity for $\ket\Phi_{RL}$ gives
\begin{equation}
 F_e\left(\mathcal R\circ\mathcal N_F,\frac{\id_L}{2}\right)
 =\frac14\sum_{i,j=0}^1
 \bra i\mathcal R\!\left(\ket i\bra j_S\otimes T_{ij}^{(F)}\right)\ket j.
 \label{eq:entfidelitysum}
\end{equation}
The diagonal terms test preservation of the logical basis states, and the off-diagonal terms test preservation of their coherence.

For achievability, apply the controlled unitary in Eq.~\eqref{eq:polarunitary} and then trace out $F$. This operation leaves the two logical basis states unchanged. Using the polar decomposition $T_F=\mathsf U_F\abs{T_F}$, the coefficient of the off-diagonal logical operator becomes
\begin{equation}
 \Tr(T_F\mathsf U_F^\dagger)=\Tr\abs{T_F}=\eta_F.
 \label{eq:achievability}
\end{equation}
The resulting channel has unchanged diagonal elements and multiplies the off-diagonal elements by $\eta_F$, giving Eq.~\eqref{eq:recoveredchannel}. Substituting these four contributions into Eq.~\eqref{eq:entfidelitysum}, we obtain
\begin{equation}
 F_e=\frac14(1+1+\eta_F+\eta_F)
    =\frac{1+\eta_F}{2}.
\end{equation}

We next show that this fidelity is optimal. Each diagonal contribution in Eq.~\eqref{eq:entfidelitysum} is at most one, since $\mathcal R$ is trace preserving. For the off-diagonal terms, the adjoint recovery map satisfies $\Tr[O\mathcal R(M)]=\Tr[\mathcal R^\dagger(O)M]$. Since $\mathcal R$ is completely positive and trace preserving, $\mathcal R^\dagger$ is unital and contracts the operator norm. Using $\abs{\Tr(KM)}\leq\norm{K}_\infty\norm{M}_1$, we obtain
\begin{equation}
 \begin{split}
 \abs{\bra0\mathcal R(\ket0\bra1_S\otimes T_F)\ket1}
     =\abs{\Tr[\mathcal R^\dagger(\ket1\bra0)
                         (\ket0\bra1_S\otimes T_F)]}
 \leq\norm{T_F}_1.
 \end{split}\label{eq:upperoffdiag}
\end{equation}
The two off-diagonal contributions are complex conjugates. Therefore,
\begin{equation}
 F_e\left(\mathcal R\circ\mathcal N_F,\frac\id2\right)
 \leq\frac{1+\eta_F}{2},\label{eq:recoveryupper}
\end{equation}
which proves Eq.~\eqref{eq:optimalrecovery} together with the recovery construction above.

The phase visibility follows from the same operator $T_F$. The difference between the coherent and dephased states is
\begin{equation}
 \rho_{SF}-\Delta_Z\rho_{SF}=
 \begin{pmatrix}
 0&\sqrt{pq}e^{-i\phi}T_F\\
 \sqrt{pq}e^{i\phi}T_F^\dagger&0
 \end{pmatrix}.\label{eq:visibilityblock}
\end{equation}
For a matrix of the form $\begin{psmallmatrix}0&A\\A^\dagger&0\end{psmallmatrix}$, the eigenvalues are the positive and negative singular values of $A$. Its trace norm is therefore $2\norm{A}_1$, proving Eq.~\eqref{eq:visibilityrecovery} with $A=\sqrt{pq}e^{-i\phi}T_F$.

To relate recovery to information in the inaccessible environment, we use $\norm{T_F}_1=\max_{U_F}\abs{\Tr(U_FT_F)}$. The maximization is over unitaries on the accessible fragment $F$. Since $\ket{r_0}$ and $\ket{r_1}$ purify $\rho_{\bar F}^{(0)}$ and $\rho_{\bar F}^{(1)}$, respectively, Uhlmann's theorem gives
\begin{equation}
 \begin{split}
 \norm{T_F}_1
 =\max_{U_F}\abs{\Tr(U_FT_F)}
 =\max_{U_F}\abs{\bra{r_1}(U_F\otimes\id_{\bar F})\ket{r_0}}
 =f(\rho_{\bar F}^{(0)},\rho_{\bar F}^{(1)}).
 \end{split}\label{eq:uhlmannproof}
\end{equation}
This proves Eq.~\eqref{eq:uhlmann}. Applying $1-f\leq D\leq\sqrt{1-f^2}$ to the two conditional states of $\bar F$, with $f=\eta_F$ and $D=D_Z(\bar F)$, gives Eqs.~\eqref{eq:fvg} and \eqref{eq:generaltradeoff}.

For the three-qubit model, the four operators defined in Eq.~\eqref{eq:tf} are
\begin{equation}
 \begin{aligned}
 T_{\varnothing}&=c_\theta,\\
 T_{E_1}&=\ket0(c_\theta\bra0+s_\theta c_\chi\bra1),\\
 T_{E_2}&=c_\theta\ket0\bra0,\\
 T_{E_1E_2}&=\ket{00}\bra{r_1}.
 \end{aligned}
 \label{eq:threeTF}
\end{equation}
Each is either a scalar or a rank-one operator. Since $\norm{\ket u\bra v}_1=\norm{u}\,\norm{v}$, their trace norms give Eqs.~\eqref{eq:etaempty}--\eqref{eq:eta12}. In particular, $\eta_{\varnothing}=\eta_{E_2}=\abs{c_\theta}$, as used in Section~5.1.

\section{Entanglement measures and phase witnesses}\label{app:entanglement}
We give the concurrence and phase-correlation calculations used in Section~6.

For a two-qubit density matrix $\rho$, define its spin-flipped matrix by $\widetilde\rho=(Y\otimes Y)\rho^*(Y\otimes Y)$, where complex conjugation is taken in the computational basis. If $\mu_1\geq\mu_2\geq\mu_3\geq\mu_4\geq0$ are the square roots of the eigenvalues of $\rho\widetilde\rho$, the concurrence is~\cite{Wootters1998}
\begin{equation}
 C(\rho)=\max(0,\mu_1-\mu_2-\mu_3-\mu_4).
 \label{eq:concurrencedef}
\end{equation}

The two-qubit reduced states in our model have rank at most two. Write such a state as $\rho=\ket{v_0}\bra{v_0}+\ket{v_1}\bra{v_1}$, where the vectors need not be normalized. The nonzero $\mu_i$ are the singular values of the $2\times2$ matrix $M_{ij}=\bra{v_i}(Y\otimes Y)\ket{v_j^*}$. This gives a short calculation without diagonalizing a $4\times4$ matrix.

For $SE_1$, trace over $E_2$ in Eq.~\eqref{eq:sequential-final-state} and use
\[
 \begin{aligned}
 \ket{v_0}&=\sqrt q\ket{00}
 +e^{i\phi}\sqrt p\,(c_\theta\ket{10}+s_\theta c_\chi\ket{11}),\\
 \ket{v_1}&=e^{i\phi}\sqrt p\,s_\theta s_\chi\ket{11}.
 \end{aligned}
\]
Their spin-flip overlap matrix is
\[
 M_{SE_1}=-\sqrt{pq}\,e^{-i\phi}s_\theta
 \begin{pmatrix}2c_\chi&s_\chi\\s_\chi&0\end{pmatrix}.
\]
For a symmetric matrix of the form $\begin{psmallmatrix}A&B\\B&0\end{psmallmatrix}$, its two singular values $\mu_\pm$ satisfy $\mu_+^2+\mu_-^2=\abs{A}^2+2\abs{B}^2$ and $\mu_+\mu_-=\abs{B}^2$. Hence $\mu_+-\mu_-=\abs{A}$. The matrix above therefore gives $C_{SE_1}=2\sqrt{pq}\abs{s_\theta c_\chi}$, as in Eq.~\eqref{eq:cs1}.

Similarly, tracing over $S$ gives $\rho_{E_1E_2}=\ket{v_0}\bra{v_0}+\ket{v_1}\bra{v_1}$ with
\[
 \ket{v_0}=\sqrt q\ket{00},\qquad
 \ket{v_1}=e^{i\phi}\sqrt p\,
 (c_\theta\ket{00}+s_\theta c_\chi\ket{10}+s_\theta s_\chi\ket{11}).
\]
The corresponding overlap matrix is
\[
 M_{E_1E_2}=-s_\theta s_\chi
 \begin{pmatrix}
 0&\sqrt{pq}\,e^{-i\phi}\\
 \sqrt{pq}\,e^{-i\phi}&2p c_\theta e^{-2i\phi}
 \end{pmatrix}.
\]
The same singular-value argument yields $C_{E_1E_2}=2p\abs{s_\theta c_\theta s_\chi}$, proving Eq.~\eqref{eq:c12}.

The remaining pair $SE_2$ is separable. Indeed, tracing out $E_1$ gives a sum of projectors onto the two unnormalized product vectors
\begin{equation}
 (\sqrt q\ket0+e^{i\phi}\sqrt p\,c_\theta\ket1)_S\ket0_{E_2}
 \quad\text{and}\quad
 e^{i\phi}\sqrt p\,s_\theta\ket1_S\ket{v_\chi}_{E_2}.
 \label{eq:se2separable}
\end{equation}
Their squared norms are $q+pc_\theta^2$ and $ps_\theta^2$, which sum to one. Thus $\rho_{SE_2}$ is a convex mixture of product states, proving $C_{SE_2}=0$ in Eq.~\eqref{eq:cs2}.

For the pure three-qubit state, the three-tangle is defined by the CKW relation~\cite{CKW2000},
\begin{equation}
 \tau_3=4\det\rho_S-C_{SE_1}^2-C_{SE_2}^2.
 \label{eq:tangledef}
\end{equation}
Substituting $\det\rho_S=pq s_\theta^2$ from Eq.~\eqref{eq:determinants} and the concurrences above gives Eqs.~\eqref{eq:tangle} and \eqref{eq:ckw}.

Finally, we derive the phase correlations used in Eqs.~\eqref{eq:xi1}--\eqref{eq:xi3}. For a Hermitian environment operator $O$, let $a=\bra{r_0}O\ket{r_1}$. In our circuit this matrix element is real for the operators considered. Expanding the cross terms in the two system branches gives
\begin{equation}
 \langle X_SO\rangle=2\sqrt{pq}\,a\cos\phi,\qquad
 \langle Y_SO\rangle=2\sqrt{pq}\,a\sin\phi.
 \label{eq:pauligeneral}
\end{equation}
From Eq.~\eqref{eq:sequential-conditional-states},
\begin{align}
 \braket{r_0}{r_1} =c_\theta,\qquad
 \bra{r_0}X_{E_1}\ket{r_1} =s_\theta c_\chi,\qquad
 \bra{r_0}X_{E_1}X_{E_2}\ket{r_1}&=s_\theta s_\chi.
 \label{eq:paulimatrixelements}
\end{align}
Squaring and adding the two phase quadratures in Eq.~\eqref{eq:pauligeneral} gives $4pq\,a^2$, and hence Eqs.~\eqref{eq:xi1}--\eqref{eq:xi3}. The resulting equalities to concurrence and three-tangle are particular to this state family and access order.

\section{Derivations for the environment chain}\label{app:chain}
% Slightly tighter display spacing keeps the derivations together without
% reducing the main-text font size.
\begingroup
\setlength{\abovedisplayskip}{8pt}
\setlength{\belowdisplayskip}{8pt}
\setlength{\abovedisplayshortskip}{4pt}
\setlength{\belowdisplayshortskip}{6pt}
This appendix gives the intermediate steps behind the chain formulas in
Section~\ref{sec:chain}. We first normalize the conditional environment state,
then trace out all but one environment qubit, and finally evaluate the
coherence retained by an accessible prefix. Throughout, the environment
starts in $\ket{0^N}$ and the controlled rotations act in the order
$U_1,U_2,\ldots,U_N$.

\paragraph{Prefix amplitudes and normalization.}
Write $c_j=\cos(\theta_j/2)$ and $s_j=\sin(\theta_j/2)$. In the $S=1$
branch, the $j$th gate acts nontrivially only on the prefix whose first
$j-1$ sites are in $\ket1$:
\begin{equation*}
\begin{aligned}
 U_j\ket{1^{j-1}0^{N-j+1}}
 &=c_j\ket{1^{j-1}0^{N-j+1}}
   +s_j\ket{1^j0^{N-j}}.
\end{aligned}
\end{equation*}
Every shorter prefix has its control in $\ket0$ and is unchanged.
Iterating the gates therefore produces
\begin{equation*}
 \ket{r_1^{(N)}}=\sum_{j=0}^N A_j\ket{1^j0^{N-j}},\qquad
 A_j=\begin{cases}
  \displaystyle\left(\prod_{i=1}^j s_i\right)c_{j+1},&0\leq j<N,\\[4pt]
  \displaystyle\prod_{i=1}^N s_i,&j=N.
 \end{cases}
\end{equation*}
The product for $j=0$ is understood to be one. This reproduces the
amplitudes in Eqs.~\eqref{eq:a0}--\eqref{eq:an}.

To see directly why they are normalized, introduce
\begin{equation*}
 B_0=1,\qquad B_j=\prod_{i=1}^j s_i^2\ (1\leq j\leq N),
 \qquad B_{N+1}=0.
\end{equation*}
For every $0\leq j<N$,
$\abs{A_j}^2=B_jc_{j+1}^2=B_j-B_{j+1}$; the same
last equality holds for $j=N$ because $\abs{A_N}^2=B_N$.
Consequently,
\begin{equation}
\begin{aligned}
 \sum_{j=0}^N\abs{A_j}^2
 &=c_1^2+s_1^2c_2^2+\cdots+\prod_{j=1}^N s_j^2\\
 &=\sum_{j=0}^N(B_j-B_{j+1})=B_0-B_{N+1}=1.
\end{aligned}
\label{eq:telescoping}
\end{equation}
The same cancellation starting at an arbitrary site $k$ gives the useful
tail identity
\begin{equation*}
 \sum_{j=k}^N\abs{A_j}^2
 =\sum_{j=k}^N(B_j-B_{j+1})
 =B_k=\prod_{i=1}^k s_i^2.
\end{equation*}
In particular, $B_k$ is the conditional probability that the excitation
has reached site $E_k$.

\paragraph{One-site reduced state and branch distinguishability.}
Consider the reduced state of $E_k$ conditional on $S=1$. A prefix with
$j\geq k$ has $E_k=1$, whereas a prefix with $j<k$ has $E_k=0$.
The two populations are therefore $B_k$ and $1-B_k$.
For an off-diagonal term to survive tracing out the other sites, its two
prefix strings must differ \emph{only} at $E_k$. This is possible only
for the adjacent prefixes of lengths $k-1$ and $k$. Thus, writing
$u_k=A_{k-1}A_k^*$, we obtain
\begin{equation*}
 \rho_{E_k}^{(1)}=
 \begin{pmatrix}1-B_k&u_k\\u_k^*&B_k\end{pmatrix},
 \qquad \rho_{E_k}^{(0)}=\ket0\bra0.
\end{equation*}
This explains both entries appearing in Eq.~\eqref{eq:chainrho}, rather
than merely quoting the partial trace.

The difference between the two conditional states is the traceless
matrix
\begin{equation*}
 \rho_{E_k}^{(0)}-\rho_{E_k}^{(1)}
 =\begin{pmatrix}B_k&-u_k\\-u_k^*&-B_k\end{pmatrix}.
\end{equation*}
Its eigenvalues are $\pm\sqrt{B_k^2+\abs{u_k}^2}$. The trace distance
is therefore
\begin{equation*}
 D_Z(E_k)=\frac12\bigl\|\rho_{E_k}^{(0)}-
 \rho_{E_k}^{(1)}\bigr\|_1
 =\sqrt{B_k^2+\abs{u_k}^2},
\end{equation*}
as stated in Eq.~\eqref{eq:chainrecord}. The $B_k$ term measures a
population difference, while $u_k$ is the contribution that can be read
through a transverse local observable.

For the physical branch probabilities $q=1-p$ and $p$, the unconditional
state is
\begin{equation*}
 \rho_{E_k}=q\rho_{E_k}^{(0)}+p\rho_{E_k}^{(1)}
 =\begin{pmatrix}1-pB_k&pu_k\\pu_k^*&pB_k\end{pmatrix}.
\end{equation*}
Since the purity deficit of a qubit is twice the determinant of its
reduced state, this immediately yields
\begin{equation*}
 P_{E_k}=2\det\rho_{E_k}
 =2pB_k(1-pB_k)-2p^2\abs{u_k}^2,
\end{equation*}
which is Eq.~\eqref{eq:chainpurity}. Similarly, the mixedness within
the $S=1$ branch is
\begin{equation*}
 \mathcal P(\rho_{E_k}^{(1)})
 =2\bigl[B_k(1-B_k)-\abs{u_k}^2\bigr].
\end{equation*}
Combining this expression with $D_Z(E_k)^2=B_k^2+\abs{u_k}^2$
gives the decomposition
\begin{equation*}
 P_{E_k}=p\,\mathcal P(\rho_{E_k}^{(1)})
       +2pq\,D_Z(E_k)^2.
\end{equation*}
Thus the purity deficit includes mixedness generated \emph{within} the
conditional environment branch, as well as the contribution associated
with distinguishing the two system branches. This is the identity used
in Eq.~\eqref{eq:chainvariance}.

\paragraph{Recovery from an accessible prefix.}
Let $F_k=E_1\cdots E_k$ be the accessible environment fragment, and
suppose the remaining sites $E_{k+1},\ldots,E_N$ are discarded. The
reference branch is $\ket{r_0^{(N)}}=\ket{0^N}$. In the cross operator
$\ket{0^N}\bra{r_1^{(N)}}$, a prefix of length $j>k$ differs from this
reference branch on at least one discarded site. Its contribution
therefore vanishes under the partial trace. Only $0\leq j\leq k$
survive, giving
\begin{equation*}
\begin{aligned}
 T_{F_k}
 &=\operatorname{Tr}_{E_{k+1}\cdots E_N}
   \bigl(\ket{0^N}\bra{r_1^{(N)}}\bigr)\\
 &=\ket{0^k}\sum_{j=0}^k A_j^*\bra{1^j0^{k-j}}.
\end{aligned}
\end{equation*}
For $k=0$, the empty tensor products are understood as scalars.
Defining the (generally unnormalized) vector
$\ket{v_k}=\sum_{j=0}^k A_j\ket{1^j0^{k-j}}$ makes the structure
transparent: $T_{F_k}=\ket{0^k}\bra{v_k}$ is a rank-one operator,
except when it vanishes. Hence
\begin{equation*}
 T_{F_k}T_{F_k}^{\dagger}
 =\left(\sum_{j=0}^k\abs{A_j}^2\right)
   \ket{0^k}\bra{0^k},
 \qquad
 \eta_{F_k}=\norm{T_{F_k}}_1
 =\sqrt{\sum_{j=0}^k\abs{A_j}^2}.
\end{equation*}
The optimal recovery and phase-visibility formulas of Section~5 then
become
\begin{equation*}
 F_Q(F_k)=\frac12\left(1+\sqrt{\sum_{j=0}^k\abs{A_j}^2}\right),
 \qquad
 V(F_k)=2\sqrt{pq}\sqrt{\sum_{j=0}^k\abs{A_j}^2}.
\end{equation*}
These are the results in Eqs.~\eqref{eq:chaineta} and
\eqref{eq:chainrecovery}. In particular, each added site contributes
one new squared amplitude:
\begin{equation*}
 w_1=\abs{A_0}^2,\qquad
 w_{k+1}=\eta_{F_k}^2-\eta_{F_{k-1}}^2
 =\abs{A_k}^2\quad(1\leq k\leq N).
\end{equation*}
The increments refer to successively larger accessible regions for the
same encoded state, not to successive times in the dynamics.

\paragraph{Pauli correlations and phase access.}
Finally, define $O_k=\prod_{j=1}^k X_{E_j}$, with $O_0=\id$.
Applied to a prefix of length $k$, this product flips precisely its
first $k$ ones and produces $\ket{0^N}$. Applied to any other prefix,
it leaves at least one bit different from the reference branch.
Therefore
\begin{equation*}
 \bra{0^N}O_k\ket{r_1^{(N)}}=A_k.
\end{equation*}
Using the full state
$\ket\Psi=\sqrt q\ket0\ket{0^N}
+e^{i\phi}\sqrt p\ket1\ket{r_1^{(N)}}$ gives
\begin{align*}
 \langle X_SO_k\rangle
 &=2\sqrt{pq}\,\operatorname{Re}(e^{i\phi}A_k),\\
 \langle Y_SO_k\rangle
 &=2\sqrt{pq}\,\operatorname{Im}(e^{i\phi}A_k).
\end{align*}
The two quadratures eliminate the dependence on the input phase when
they are squared and added:
\begin{equation*}
 \langle X_SO_k\rangle^2+\langle Y_SO_k\rangle^2
 =4pq\abs{A_k}^2=4pq\,w_{k+1}.
\end{equation*}
This is Eq.~\eqref{eq:chainwitness}. It connects a measurable joint
correlation to the recovery increment for the specified prefix access
order. It is not a general measure of multipartite entanglement.
\endgroup

%==========================================================

\bibliography{references}
\bibliographystyle{unsrt}
\end{document}